\documentclass[aps,pra,twocolumn,floatfix]{revtex4-1}
\usepackage{graphics}% Include figure files
\usepackage{bm}% bold matha
\usepackage[caption=false,labelformat=simple]{subfig}% For subfigure
\usepackage[dvips]{graphicx}
\usepackage{amsmath}
\usepackage{amssymb}
\usepackage{blkarray}
\usepackage[dvipsnames]{xcolor}
\usepackage[colorlinks=true,linkcolor=blue]{hyperref}

\renewcommand{\d}{{d}}
\begin{document}
\draft

\title{Feshbach resonances and molecule formation \\
in ultracold mixtures of Rb and Yb($^3$P) atoms}

\author{Bijit Mukherjee}
\affiliation{Joint Quantum Centre (JQC) Durham-Newcastle, Department of
Chemistry, Durham University, South Road, Durham, DH1 3LE, United Kingdom.}
\author{Matthew D. Frye}
\affiliation{Joint Quantum Centre (JQC) Durham-Newcastle, Department of
Chemistry, Durham University, South Road, Durham, DH1 3LE, United Kingdom.}
\author{Jeremy M. Hutson}
\email{j.m.hutson@durham.ac.uk} \affiliation{Joint Quantum Centre (JQC)
Durham-Newcastle, Department of Chemistry, Durham University, South Road,
Durham, DH1 3LE, United Kingdom.}

\date{\today}

\begin{abstract}
We have investigated magnetically tunable Feshbach resonances in ultracold collisions of Rb with Yb in its metastable $^3$P$_2$ and $^3$P$_0$ states, using coupled-channel scattering and bound-state calculations. For the $^3$P$_2$ state, we find sharp resonances when both atoms are in their lowest Zeeman sublevels. However, these resonances are decayed by inelastic processes that produce Yb atoms in $^3$P$_1$ and $^3$P$_0$ states. The molecules that might be produced by magnetoassociation at the  $^3$P$_2$ thresholds can decay by similar pathways and would have lifetimes no more than a few microseconds. For the $^3$P$_0$ state, by contrast, there are resonances that are promising for magnetoassociation. There are resonances due to both rotating and non-rotating molecular states that are significantly stronger than the analogous resonances for Yb($^1$S). The ones due to rotating states are denser in magnetic field; in contrast to Yb($^1$S), they exist even for bosonic isotopes of Yb($^3$P$_0$).
\end{abstract}

\maketitle

\section{Introduction}

There is great interest in forming ultracold molecules containing an alkali-metal atom and a closed-shell atom such as Sr or Yb. Such molecules have $^2\Sigma$ electronic ground states with both electric and magnetic dipole moments. Potential applications include the study of lattice spin models in many-body physics \cite{Micheli:2006} and searches for the electric dipole moment of the electron
\cite{Meyer:2009}.

Ultracold molecules such as the alkali-metal diatomics are commonly formed by magnetoassociation, in which pairs of atoms are converted to molecules by tuning a magnetic field across a zero-energy Feshbach resonance. Such resonances exist when high-lying molecular bound states cross an atomic threshold, and are coupled to it to form an avoided crossing. In the alkali-metal pairs, there are two main coupling mechanisms that can cause Feshbach resonances. First, the existence of singlet and triplet spin states of the molecule, with different potential-energy curves, provides a coupling between different atomic hyperfine states. This coupling can cause resonances in s-wave scattering due to s-wave bound states, with relative angular momentum $L=0$. Secondly, magnetic dipole-dipole coupling between unpaired electrons on the two atoms can cause resonances in s-wave scattering due to d-wave and higher-wave bound states with $L\ge2$. Both types of resonance have been extensively used, both to tune scattering lengths and interaction strengths and to form molecules by magnetoassociation.

Neither of the mechanisms that dominate for alkali-metal pairs exists for mixtures of alkali metals with closed-shell atoms. A closed-shell atom in a $^1$S state interacts with an alkali-metal atom in a $^2$S state to form only a single molecular state, of $^2\Sigma$ symmetry. This provides no couplings between hyperfine states. In addition, a closed-shell atom has no magnetic dipole (except possibly a nuclear dipole) to provide a dipole-dipole interaction. The only resonances that exist for mixtures of alkali-metal and closed-shell atoms are due to weaker interactions, and are much narrower. In particular, the hyperfine coupling of the alkali-metal atom is modified by the presence of the closed-shell atom, and this produces a weak coupling that can produce resonances due to s-wave bound states \cite{Zuchowski:RbSr:2010, Brue:AlkYb:2013}. This can occur even for spin-zero (bosonic) isotopes of the closed-shell atom. In addition, if the closed-shell atom has non-zero nuclear spin, there are resonances due to its scalar and tensor interactions with the electron spin of the alkali-metal atom \cite{Brue:LiYb:2012, Yang:CsYb:2019}.

Much experimental work has been devoted to locating and observing Feshbach resonances in these systems \cite{Nemitz:2009, Baumer:2011, Muenchow:2011, Borkowski:2013, Ivanov:2011, Hansen:2011, Roy:2016, Green:2019, Guttridge:2017, Guttridge:1p:2018, Guttridge:2p:2018, Barbe:RbSr:2018}. Resonances have now been observed for both bosonic and fermionic isotopes of Sr interacting with Rb \cite{Barbe:RbSr:2018}, for fermionic $^{173}$Yb interacting with $^6$Li \cite{Green:LiYb-res:2020} and for $^{173}$Yb interacting with Cs \cite{Guttridge:CsYb-res:2022}. However, the resonances are very narrow. In addition, the ones for bosonic isotopes of Sr and Yb are very sparse in magnetic field \cite{Zuchowski:RbSr:2010, Brue:AlkYb:2013, Muenchow:thesis:2012}, and for many isotopic combinations exist only at magnetic fields that are hard to achieve in experiments. Attempts to form ultracold molecules by magnetoassociation at these resonances have so far been unsuccessful.

A possible alternative is to use atoms in excited states. In particular, ultracold samples of Sr and Yb can be prepared in the metastable $^3$P$_2$ and $^3$P$_0$ states. An atom in a $^3$P state combines with one in a $^2$S state to form multiple molecular electronic states, so broader Feshbach resonances might be expected. However, molecules formed at the $^3$P$_2$ threshold have finite lifetimes, due to predissociation by spin recoupling to form atoms in the lower $^3$P$_1$ and $^3$P$_0$ states. Previous work on Li+Yb($^3$P$_2$) has shown that the resulting resonances are strongly decayed by these inelastic processes \cite{Gonzalez-Martinez:LiYb:2013, Khramov:2014, Dowd:2015}, and are unlikely to be suitable for molecule formation. It is not immediately clear whether this will remain true for heavier alkali-metal atoms, since inelastic processes that release substantial kinetic energy are often suppressed when the reduced mass is large \cite{Beswick:1978, Ewing:1982}.

The purpose of the present work is to investigate Feshbach resonances for Yb($^3$P$_j$) interacting with Rb($^2$S). We find that resonances for Yb($^3$P$_2$) remain significantly decayed, and even in favorable cases the molecules formed at them predissociate on a microsecond timescale. However, the $^3$P$_0$ threshold is more promising. Molecules formed at this threshold cannot decay by spin recoupling. The predicted resonances are narrow, because they rely on indirect couplings between the bound state and the threshold, but they can be substantially wider than for Yb($^1$S).

The structure of the paper is as follows. Section \ref{sec:theory} describes the theoretical methods we use, including the interaction potential for Yb($^3$P$_j$) interacting with Rb($^2$S) and the specific form of the coupled-channel equations. Section III describes results for Feshbach resonances at the Yb($^3$P$_2$) thresholds, including their dependence on the interaction potential, and discusses the lifetimes of molecules that would be produced by magnetoassociation. Section IV describes results for Feshbach resonances at the Yb($^3$P$_0$) thresholds. Section V presents conclusions and perspectives.

\section{Theory} \label{sec:theory}

\subsection{Coupled-channel methods} \label{sec:cc}

We carry out coupled-channel scattering and bound-state calculations. The total wavefunction is expanded
\begin{equation} \Psi(R,\xi)
=R^{-1}\sum_j\Phi_j(\xi)\psi_{j}(R), \label{eqexp}
\end{equation}
where $R$ is the internuclear distance and the functions $\Phi_j(\xi)$ form a complete orthonormal basis set for motion in all other coordinates, collectively labeled $\xi$. For interaction of Rb($^2$S) with Yb($^3$P), $\xi$ includes the electron and nuclear spins on both atoms, the orbital angular momentum of Yb, and the relative angular momentum $L$. The factor $R^{-1}$
serves to simplify the form of the radial kinetic energy operator. The wavefunction in each {\em channel} $j$ is described by a radial \emph{channel function} $\psi_{j}(R)$.

The Hamiltonian of the interacting pair is
\begin{equation}
\hat H=-\frac{\hbar^2}{2\mu}R^{-1}\frac{\d^2\ }{\d R^2}R
+ \frac{\hbar^2 \hat L^2}{2\mu R^2} + \hat{H}_\textrm{Rb} + \hat{H}_\textrm{Yb} + \hat{V}(R,\xi).
\label{eqh}
\end{equation}
Here $\hat{H}_\textrm{Rb}$ and $\hat{H}_\textrm{Yb}$ are the Hamiltonians of the isolated Rb and Yb atoms, and depend on $\xi$ but not $R$, and $\hat{V}(R,\xi)$ is an interaction operator described below. The operator $\hbar^2 \hat L^2/2\mu R^2$ is the centrifugal term that describes the end-over-end rotational energy of the interacting pair.

Substituting the expansion (\ref{eqexp}) into the total Schr\"odinger equation, and projecting onto a basis function $\Phi_i(\xi)$, produces a set of coupled differential equations for the
channel functions $\psi_{i}(R)$,
\begin{equation}\frac{\d^2\psi_{i}}{\d R^2}
=\sum_j\left[W_{ij}(R)-{\cal E}\delta_{ij}\right]\psi_{j}(R), \label{eq:se-invlen}
\end{equation}
where $\delta_{ij}$ is the Kronecker delta, ${\cal E}=2\mu E/\hbar^2$, $E$ is
the total energy, and
\begin{align}
W_{ij}(R) = \frac{2\mu}{\hbar^2}\int \Phi_i^*&(\xi) \Bigg[\frac{\hbar^2 \hat L^2}{2\mu R^2}
+ \hat{H}_\textrm{Rb} + \hat{H}_\textrm{Yb} \nonumber\\
& + \hat{V}(R,\xi) \Bigg] \Phi_j(\xi)\,\d\xi. \label{eqWij}
\end{align}
The different equations are coupled by the off-diagonal terms $W_{ij}(R)$ with $i\ne j$.

The atomic Hamiltonian of Rb($^2$S) is
\begin{equation}
\hat{H}_\textrm{Rb} = \zeta_\textrm{Rb} \hat{i}_\textrm{Rb}\cdot\hat{s}_\textrm{Rb}
+ \left( g_{s,\textrm{Rb}} \hat{s}_{\textrm{Rb},z} + g_i\hat{i}_{\textrm{Rb},z} \right) \mu_\textrm{B} B,
\end{equation}
where $\hat{s}_\textrm{Rb}$ and $\hat{i}_\textrm{Rb}$ are vector operators for the electron and nuclear spin, $\hat{s}_{\textrm{Rb},z}$ and $\hat{i}_{\textrm{Rb},z}$ are their components along the $z$ axis defined by the magnetic field, $\zeta_\textrm{Rb}$ is the hyperfine coupling constant, and $g_{s,\textrm{Rb}}$ and $g_i$ are the g-factors for the electron and nuclear spins \footnote{In writing basis sets for pairs of atoms, it is necessary to distinguish between quantum numbers for the individual atoms and those for the pair. We adopt the widely used convention of using lower-case letters for the individual atoms and upper-case letters for the pair. For example, we use $S$ for the resultant of $s_\textrm{Rb}$ and $s_\textrm{Yb}$.}.
The atomic Hamiltonian of Yb($^3$P), neglecting any nuclear spin, is
\begin{equation}
\hat{H}_\textrm{Yb} = a_\textrm{Yb} \hat{l}_\textrm{Yb}\cdot\hat{s}_\textrm{Yb} + a_1\delta_{j1}
+ \left( \hat{l}_{\textrm{Yb},z} + g_{s,\textrm{Yb}}\hat{s}_{\textrm{Yb},z} \right) \mu_\textrm{B} B,
\end{equation}
where $\hat{l}_\textrm{Yb}$ and $\hat{s}_\textrm{Yb}$ are vector operators for the electron orbital angular momentum and spin, $\hat{l}_{\textrm{Yb},z}$ and $\hat{s}_{\textrm{Yb},z}$ are their components along the $z$ axis. The atomic spin-orbit coupling constant $a_\textrm{Yb}/hc = 807.3163$ cm$^{-1}$ is chosen to reproduce the splitting between the $^3$P$_0$ and $^3$P$_2$ states and $a_1/hc = -103.7483$ cm$^{-1}$ shifts the $^3$P$_1$ state down in energy to account for the effects of $jj$ coupling. $g_{s,\textrm{Yb}}$ is the g-factor for the electron spin; here we use the free-electron value, which is very slightly different from $g_{s,\textrm{Rb}}$.

In the present work, we solve the coupled-channel equations subject to both scattering and bound-state boundary conditions. Scattering calculations are performed with the \textsc{molscat} package \cite{molscat:2019, mbf-github:2020}, with purpose-written plug-in routines to implement the basis sets and interaction operators described here. Such calculations produce the scattering matrix $\boldsymbol{S}$, for a single value of the collision energy and magnetic field each time. The complex s-wave scattering length $a$ is obtained from the diagonal element of $\boldsymbol{S}$ in the incoming channel, $S_{00}$, using the identity \cite{Hutson:res:2007}
\begin{equation}
a(k_0) = \frac{1}{ik_0} \left(\frac{1-S_{00}(k_0)}{1+S_{00}(k_0)}\right),
\end{equation}
where $k_0$ is the incoming wavenumber, related to the collision energy $E_\textrm{coll}$ by $E_\textrm{coll}=\hbar^2k_0^2/(2\mu)$. The scattering length $a(k_0)$ becomes constant at sufficiently low $E_\textrm{coll}$, and in the present work calculations are performed at $E_\textrm{coll}/k_\textrm{B} = 100$ nK.

Coupled-channel bound-state calculations are performed using the packages \textsc{bound} and \textsc{field} \cite{bound+field:2019, mbf-github:2020}, which converge upon bound-state energies at fixed field, or bound-state fields at fixed energy, respectively. These packages use the same plug-in routines for basis sets and interaction operators as \textsc{molscat}. The methods used are described in Ref.\ \cite{Hutson:CPC:1994}.

\textsc{molscat}, \textsc{bound} and \textsc{field} implement many different propagators for solving the coupled-channel equations. In the present work, the coupled equations for both scattering and bound-state calculations are solved using the fixed-step symplectic propagator of Manolopoulos and Gray \cite{MG:symplectic:1995} from $R_\textrm{min}=3$~\AA\ to $R_\textrm{mid}=18$~\AA, with an interval size of 0.001~\AA, and the variable-step Airy propagator of Alexander and Manolopoulos \cite{Alexander:1987} from $R_\textrm{mid}$ to $R_\textrm{max}$. We used $R_\textrm{max}=2000$~\AA\ and 100~\AA\ for scattering and bound-state calculations, respectively.

\subsection{Basis sets} \label{sec:basis}

The many different angular momenta may be coupled together in several different ways, and different coupling schemes are useful when discussing different aspects of the problem. The separated atoms are best represented by quantum numbers $(s_\textrm{Rb},i_\textrm{Rb})f,m_f$ and $(l_\textrm{Yb},s_\textrm{Yb})j,m_j$, where the notation $(a,b)c$ indicates that $c$ is the resultant of $a$ and $b$ and $m_c$ is the projection of $c$ onto the $z$ axis. Here $f$ and $j$ are not strictly conserved quantum numbers in a magnetic field, but serve to identify states by adiabatic correlation as a function of field.

At shorter range, the couplings change in important ways. $l_\textrm{Yb}$ decouples from $s_\textrm{Yb}$, and in the absence of spin-orbit coupling would quantize along the internuclear axis with body-fixed projection $\lambda$. $s_\textrm{Rb}$ decouples from $i_\textrm{Rb}$ and couples to $s_\textrm{Yb}$ with resultant $S$, which can be considered either with projection $M_S$ along $z$ or with projection $\Sigma$ onto the molecular axis. Finally, molecular spin-orbit coupling mixes states with different $\lambda$, $S$, and $\Sigma$ but the same $\Omega=\lambda+\Sigma$. Despite the mixing, $S$ and $\lambda$ have some useful meaning around the potential minimum and the electronic states are conventionally described with labels such as $^2\Pi_{3/2}$, indicating $2S+1=2$ (so $S=\frac{1}{2}$), $|\lambda|=1$ and $|\Omega|=\frac{3}{2}$. $\Omega$ is conserved with respect to the \emph{electronic} parts of the Hamiltonian, but different values of $\Omega$ are nevertheless mixed by Coriolis terms arising from the centrifugal operator $\hbar^2\hat{L}^2/2\mu R^2$ and the hyperfine Hamiltonian.

To carry out coupled-channel calculations, we need a basis set that spans the complete space, including relative rotation and nuclear spin. We do not require a basis set where $\hat{H}_\textrm{Rb}$ and $\hat{H}_\textrm{Yb}$ are diagonal, because \textsc{molscat} transforms the solutions of the coupled equations into an asymptotically diagonal basis set before applying scattering boundary conditions. We therefore choose to use the basis set
\begin{equation}
|s_\textrm{Rb},m_{s,\textrm{Rb}}\rangle
|i_\textrm{Rb},m_{i,\textrm{Rb}}\rangle
|(l_\textrm{Yb},s_\textrm{Yb})j,m_j\rangle
|L,M_L\rangle
\label{eq:basis}
\end{equation}
The only conserved quantities in a magnetic field are $M_\textrm{tot} = m_{s,\textrm{Rb}} + m_{i,\textrm{Rb}} + m_j + M_L$ and parity $(-1)^{L+l_\textrm{Yb}}$. We take advantage of this to perform calculations for each $M_\textrm{tot}$ and parity separately. In each calculation, we include all basis functions of the required $M_\textrm{tot}$ and parity for $s_\textrm{Rb}=\frac{1}{2}$, $i_\textrm{Rb}=\frac{3}{2}$ for $^{87}$Rb, $l_\textrm{Yb}=1$, $s_\textrm{Yb}=1$, subject to the limitation $L\le L_\textrm{max}$.
%For special purposes, described below, we can choose to limit $j$ to specific values.

\subsection{The interaction operator} \label{sec:potential}

If spin-orbit coupling is neglected, a Yb atom in its $^3$P state interacts with an alkali-metal atom in a $^2$S state to form four molecular electronic states, $^2\Sigma$, $^2\Pi$, $^4\Sigma$ and $^4\Pi$. There are additional molecular states arising from $^2$P and higher states of the alkali-metal atom, from the $^1$P state of Yb, and even from ion-pair states. In the present work we make the approximation that only the four electronic states arising from $^3$P+$^2$S contribute significantly and that the effects of other states can be included through perturbative effects on the potential curves.

Spin-orbit coupling splits the $^3$P state of Yb into three fine-structure components $^3$P$_j$, where $j=0$, 1, 2 is the total atomic electronic angular momentum. At short range it splits the $^2\Pi$ molecular state into components with $|\Omega|=\frac{1}{2}$ and $\frac{3}{2}$ and the $^4\Pi$ state into four components, two with $|\Omega|=\frac{1}{2}$ and others with $|\Omega|=\frac{3}{2}$ and $\frac{5}{2}$. Here $\Omega$ is the projection of all the electronic orbital and spin angular momenta onto the molecular axis. However, only $|\Omega|$ is conserved in the electronic Hamiltonian, and states with $\Omega=+\frac{1}{2}$ and $-\frac{1}{2}$ can mix.

Shundalau and Minko \cite{Shundalau:2017} have carried out electronic structure calculations on the ground and low-lying excited states of RbYb, using complete-active-space self-consistent-field (CASSCF) methods with perturbative corrections. They presented Born-Oppenheimer potential curves for all the relevant electronic states, including spin-orbit coupling, as a function of internuclear distance $R$. However, the Born-Oppenheimer curves are not sufficient to carry out coupled-channel calculations of the bound states and scattering, as are needed to predict ultracold experiments. We need both the potential energy curves and the couplings between them, in either an adiabatic or (preferably) a diabatic representation.

\begin{figure*}[htp]
	\subfloat[]{
		\includegraphics[width=0.45\textwidth]{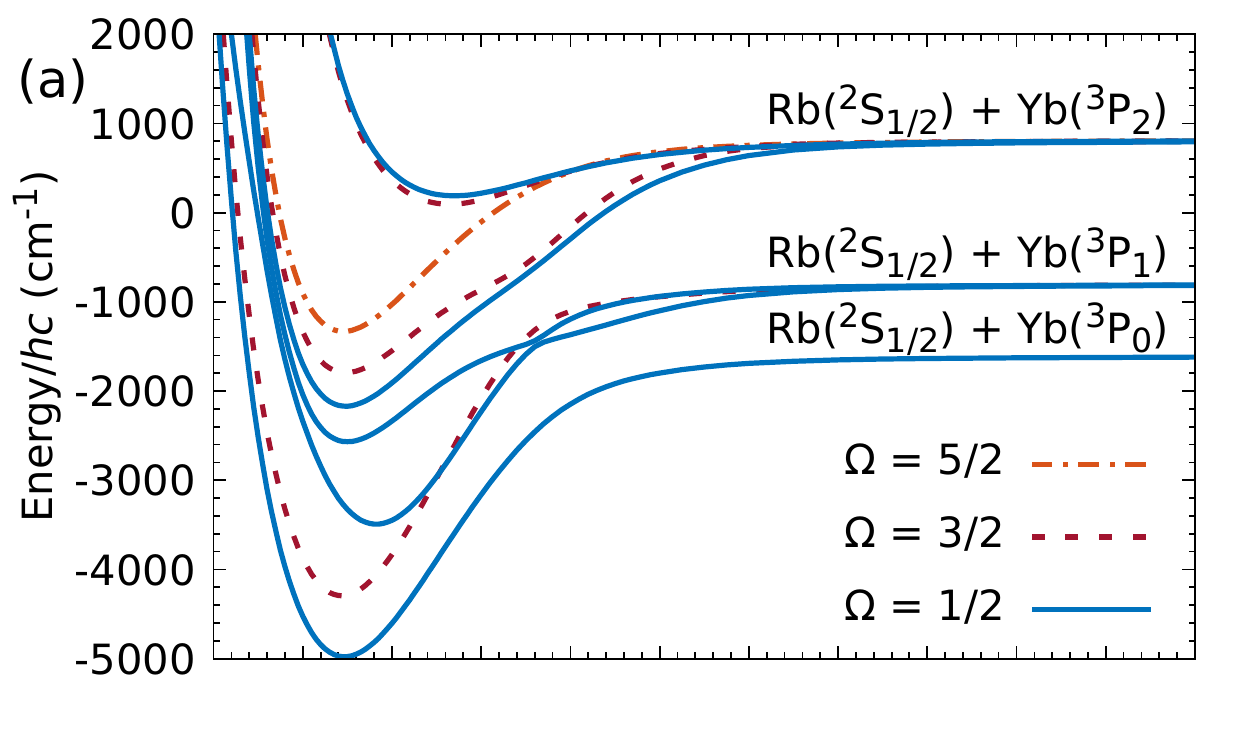}
	}
	\subfloat[]{
		\includegraphics[width=0.45\textwidth]{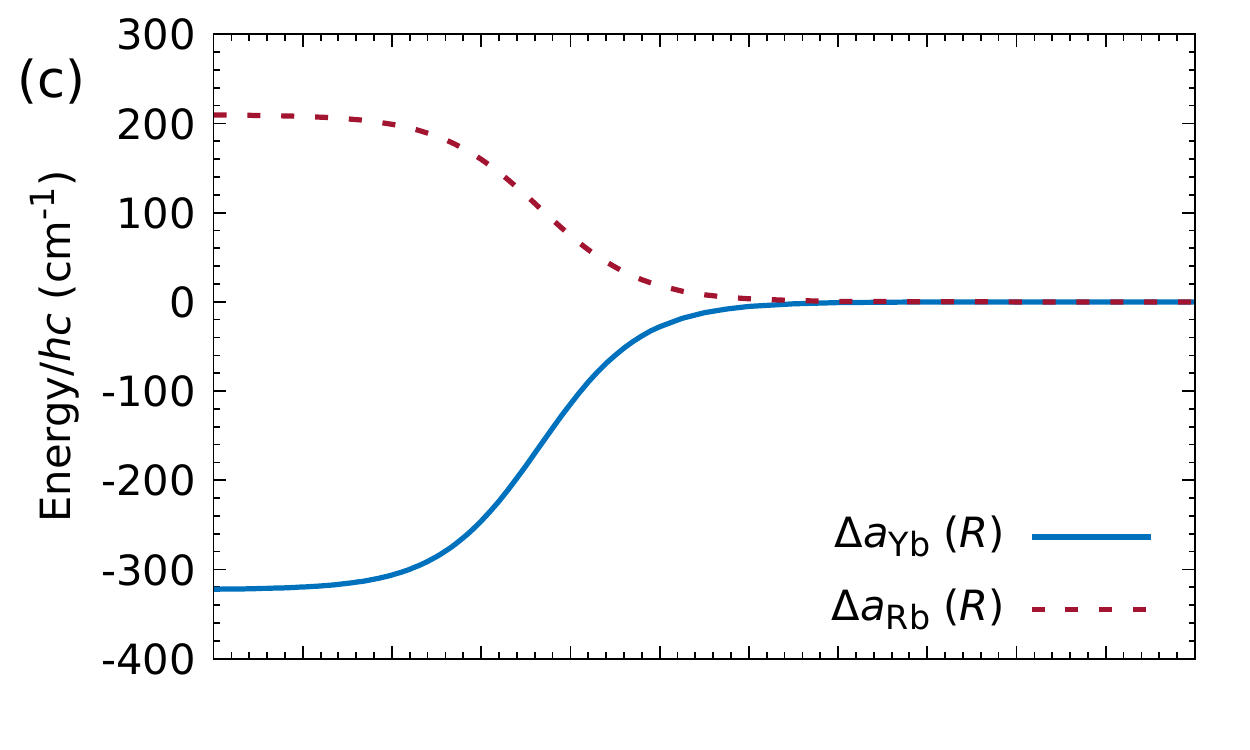}
	}
    %%%%%%%%%%%%%%%%%%%%%%%%%%%%%%%%%%%%second row
    \vspace{-1.2 cm}
	\subfloat[]{
		\includegraphics[width=0.45\textwidth]{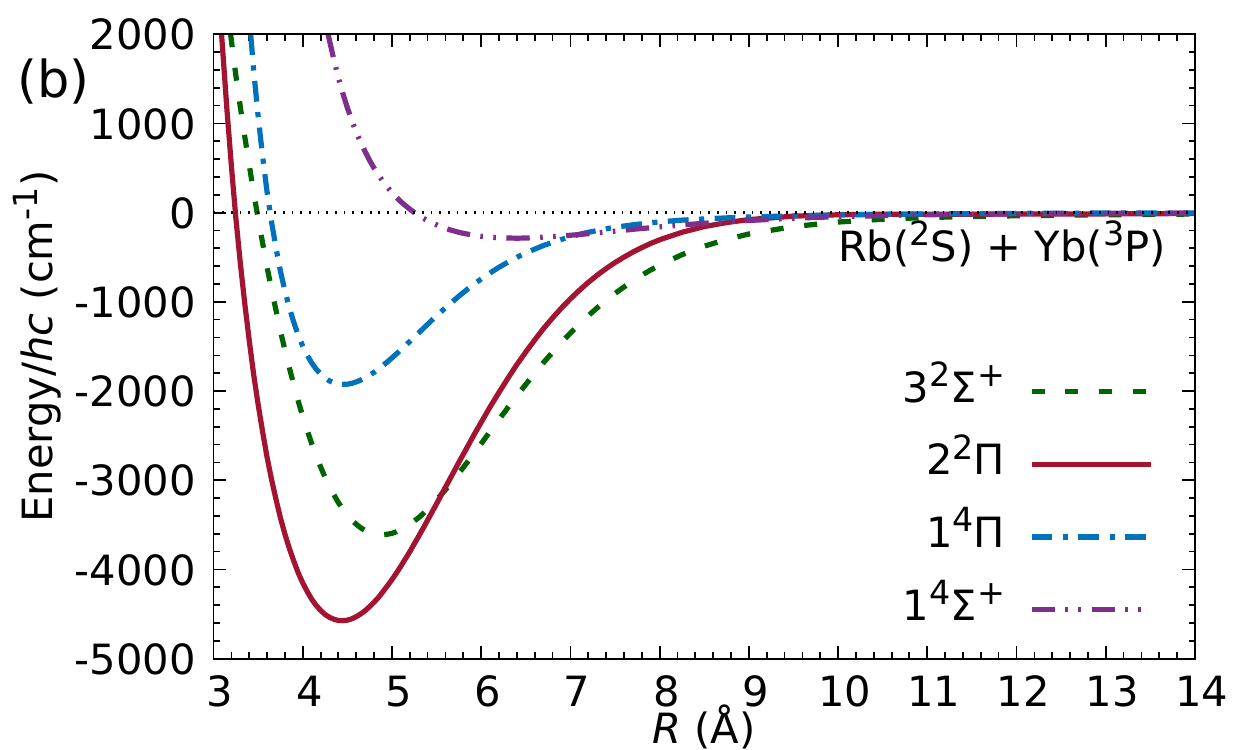}
	}
	\subfloat[]{
		\includegraphics[width=0.45\textwidth]{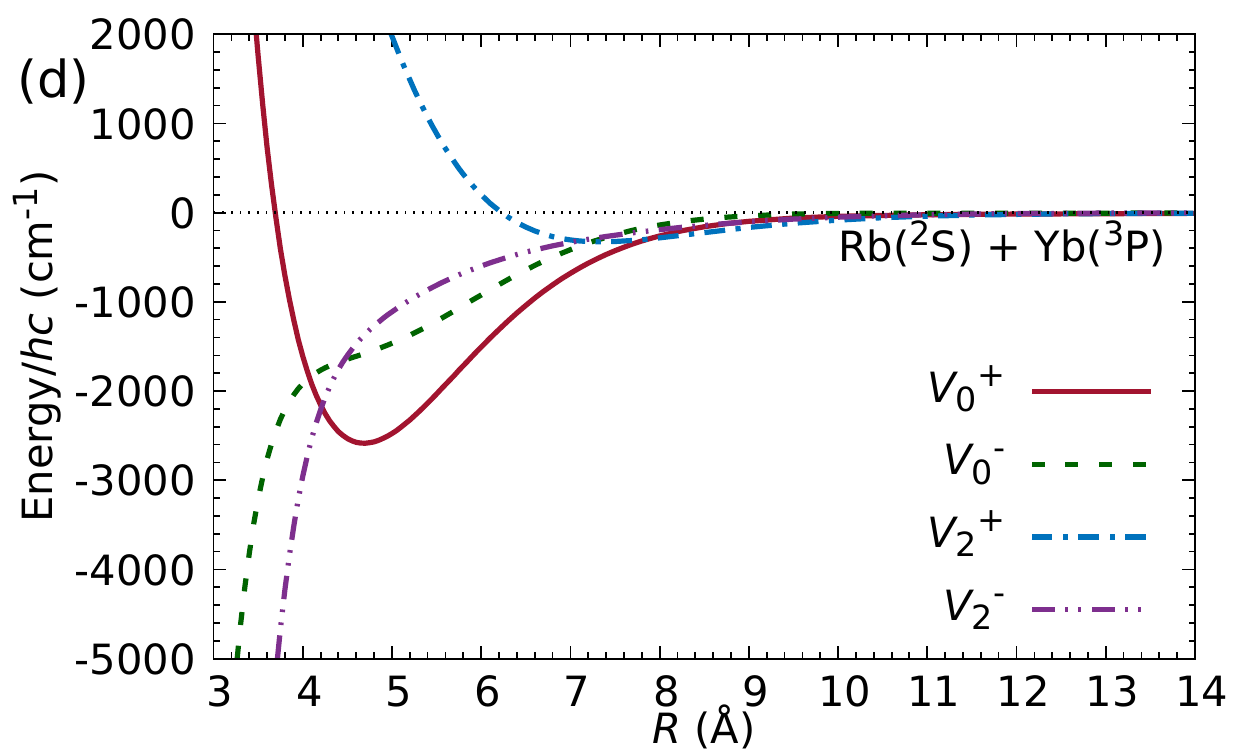}
	}
    \vspace{-0.5 cm}
    \caption{(a) Adiabatic potential energy curves, including spin-orbit coupling; (b) spin-orbit-free potential curves; (c) spin-orbit coupling 	
    functions; (d) spin-averaged and spin-difference potential curves.
}%
\label{fig:pot-curves}
\end{figure*}

It is therefore necessary to model the CASSCF potential curves in a representation that provides both diagonal and off-diagonal matrix elements. We initially hoped that we could fit the curves of ref.\ \cite{Shundalau:2017} using a set of diabatic curves for the spin-orbit-free $^2\Sigma$, $^2\Pi$, $^4\Sigma$ and $^4\Pi$ states, together with an $R$-independent spin-orbit coupling operator for the Yb atom, to provide a representation analogous to that used for Li+Yb($^3$P) \cite{Gonzalez-Martinez:LiYb:2013}. However, this representation proved unsatisfactory for Rb+Yb($^3$P). To reproduce the CASSCF curves, it was necessary to use a more complicated $R$-dependent spin-orbit coupling operator. We choose the form
\begin{equation}
\hat{V}_\textrm{so}(R,\xi) = \Delta a_\textrm{Yb}(R) \hat{l}_\textrm{Yb}\cdot\hat{s}_\textrm{Yb}
+ \Delta a_\textrm{Rb}(R) \hat{l}_\textrm{Yb}\cdot\hat{s}_\textrm{Rb},
\end{equation}
where the second term accounts for the interaction between the spin originally on Rb with the orbital angular momentum.
The potential curves for the spin-orbit-free states are described by Hulburt-Hirschfelder curves \cite{Hulburt:1941}, supplemented by damped dispersion terms at long range. The spin-orbit-free curves, spin-orbit functions and the resulting adiabatic curves including spin-orbit coupling are shown in Fig.\ \ref{fig:pot-curves}. The fitted parameters for the potential curves and spin-orbit matrix are described in Appendix A.

In addition to couplings due to electrostatic and spin-orbit interactions, there is a magnetic dipole-dipole interaction between the electron spin on the Rb and the orbital and spin angular momenta on the Yb atom \cite{Gonzalez-Martinez:H+F:2013}.
However, Rb($^2$S) + Yb($^3$P) contrasts with the alkali-metal pairs, where the magnetic dipole-dipole interaction, although weak, is the dominant anisotropic term capable of mixing channels of different $L$. In the present system, there are far stronger terms off-diagonal in $L$ that arise from electrostatic interactions, so that the dipole-dipole term is less important in comparison.

The complete interaction operator may be written
\begin{equation}
\hat{V}(R) = \sum_{S,\lambda} V^S_\lambda(R) \hat{\cal V}^S_\lambda + \hat{V}_\textrm{so}(R,\xi) + \hat{V}^\textrm{d}
\label{eq:complete-int}
\end{equation}
where the operator $\hat{\cal V}^S_\lambda = | ^{2S+1}\lambda \rangle \langle ^{2S+1}\lambda |$ projects onto a single orbital projection $\lambda$ and spin multiplicity $2S+1$.

The matrix elements of the spin-free interaction potentials, $\hat{l}_\textrm{Yb}\cdot\hat{s}_\textrm{Yb}$, the Zeeman interaction and the magnetic dipole-dipole interaction in the basis set (\ref{eq:basis}) have been given previously \cite{Gonzalez-Martinez:H+F:2013}. The matrix elements of $\hat{l}_\textrm{Yb}\cdot\hat{s}_\textrm{Rb}$ are given in Appendix B.

\subsection{Selection rules}

It is helpful to think of the terms in the interaction operator in terms of their spherical tensor character, since this determines the selection rules that govern their matrix elements. Any term in the interaction must be unchanged by overall rotations of the system in space, so must be scalar (rank 0) in the total angular momentum $F$. However, it may have internal structure, and non-zero rank with respect to some of the component angular momenta.

To describe the tensor character, we represent the potential operators for $\Sigma$ and $\Pi$ states in terms of isotropic and anisotropic components, $V_0$ and $V_2$, for each total spin, $S=1/2$ (doublet) and $S=3/2$ (quartet) \cite{DUBERNET:open:1994},
\begin{align}
\label{eq:v02}
V_0^S(R) &= \frac{1}{3} \left( V_\Sigma^S(R) + 2V_\Pi^S(R) \right) \\
V_2^S(R) &= \frac{5}{3} \left( V_\Sigma^S(R) -  V_\Pi^S(R) \right).
\end{align}
We then define averages and differences of the potentials for the two spins,
\begin{equation}
V^\pm_\kappa(R) = \frac{1}{2} \left( V_\kappa^{1/2}(R) \pm V_\kappa^{3/2}(R) \right).
\end{equation}
The resulting potential curves are shown in Fig.\ \ref{fig:pot-curves}(d).
We represent the corresponding operators as $\hat{\cal V}^\pm_\kappa$, so that the complete interaction operator is
\begin{equation}
\hat{V}(R) = \sum_{\pm,\kappa} V^\pm_\kappa(R) \hat{\cal V}^\pm_\kappa + \hat{V}_\textrm{so}(R,\xi) + \hat{V}^\textrm{d}.
\label{eq:vpm}
\end{equation}
This is equivalent to Eq.\ (\ref{eq:complete-int}), but written in terms of the average and difference potentials of Eqs.\ (\ref{eq:v02}) to (\ref{eq:vpm}).

The spherical tensor character of the operators is most simply expressed in a representation where $s_\textrm{Rb}$ and $s_\textrm{Yb}$ couple to give resultant $S$, and $l_\textrm{Yb}$ and $L$ couple to give resultant $N$, which is the spin-free angular momentum. $N$ and $S$ then couple to give $J$, the total angular momentum excluding nuclear spin. We represent the tensor rank by a superscript integer for each component angular momentum.

In this representation, $\hat{\cal V}^+_0$ is a scalar operator in all the component angular momenta, whose matrix representation is a unit matrix in any basis set.
$\hat{\cal V}^-_0$ has tensor character $\big( \mathbf{s}_\textrm{Rb}^{(1)} \otimes \mathbf{s}_\textrm{Yb}^{(1)} \big)^{(0)}$, so has matrix elements that can change $m_{s,\textrm{Rb}}$ and $m_{s,\textrm{Yb}}$ by 1 in opposite directions, while conserving their sum. It is worth noting that exactly the same tensor character and selection rules apply to the difference between the singlet and triplet potentials in the alkali-metal pairs, where $V^-_0(R) = \frac{1}{2} \big( V^0(R) - V^1(R) \big)$. $\hat{\cal V}^+_2$ has tensor character $\big( \mathbf{l}_\textrm{Yb}^{(2)} \otimes \mathbf{L}^{(2)} \big)^{(0)}$, so has matrix elements that can change $m_{l,\textrm{Yb}}$ and $m_L$ by 1 or 2 in opposite directions, while conserving their sum. $\hat{\cal V}^-_2$ has tensor character $\big( \mathbf{s}_\textrm{Rb}^{(1)} \otimes \mathbf{s}_\textrm{Yb}^{(1)} \big)^{(0)} \otimes \big( \mathbf{l}_\textrm{Yb}^{(2)} \otimes \mathbf{L}^{(2)} \big)^{(0)}$, so has matrix elements with selection rules that are a combination of those for $\hat{\cal V}^-_0$ and $\hat{\cal V}^+_2$.
The atomic spin-orbit operators $\hat{l}_\textrm{Yb} \cdot \hat{s}_\textrm{Yb}$ and $\hat{l}_\textrm{Yb} \cdot \hat{s}_\textrm{Rb}$ have tensor characters $\big( \mathbf{l}_\textrm{Yb}^{(1)} \otimes \mathbf{s}_\textrm{Yb}^{(1)} \big)^{(0)}$ and $\big( \mathbf{l}_\textrm{Yb}^{(1)} \otimes \mathbf{s}_\textrm{Rb}^{(1)} \big)^{(0)}$, respectively. The magnetic dipole-dipole operator has tensor character $\big[ \big( \mathbf{s}_\textrm{Rb}^{(1)} \otimes \mathbf{s}_\textrm{Yb}^{(1)} \big)^{(2)} \otimes \mathbf{L}^{(2)} \big]^{(0)}$.

In the discussion below, we often need matrix elements in basis sets that are coupled in a different sequence to those used for the operators above. These are obtained using standard equations for the recoupling of tensor operators \cite{brink}. For example, in a representation labeled by $j$ and $m_j$ instead of $m_{l,\textrm{Yb}}$ and $m_{s,\textrm{Yb}}$, the tensor character of $\hat{\cal V}^-_0$ may be expanded as $\big[ \mathbf{s}_\textrm{Rb}^{(1)} \otimes \big( \mathbf{l}_\textrm{Yb}^{(0)} \otimes \mathbf{s}_\textrm{Yb}^{(1)} \big)^{(1)} \big]^{(0)}$, so it has matrix elements that can change $j$ by 0 or 1 and $m_j$ and $m_{s,\textrm{Rb}}$ by 1 in opposite directions, while conserving their sum. In such a basis set the selection rules may be summarized
\begin{equation} \hbox{
\begin{blockarray}{cccc}
\quad $j$\qquad & 2              & 1                          & 0 \\
\begin{block}{c(ccc)}
  2 & $V_0^+,V_0^-,V_2^+,V_2^-$ & $V_0^-,V_2^+,V_2^-$        & $V_2^+,V_2^-$  \\
  1 & $V_0^-,V_2^+,V_2^-$       & $V_0^+,V_0^-,V_2^+,V_2^-$  & $V_0^-$  \\
  0 & $V_2^+,V_2^-$             & $V_0^-$                    & $V_0^+$  \\
\end{block}
\end{blockarray}}
\end{equation}
Operators with superscript $-$ can change $m_j$ and $m_{s,\textrm{Rb}}$ (and thus $f$ and $m_f$), while those with $\kappa=2$ can change $L$ and $M_L$ as well as $j$ and $m_j$. The Yb spin-orbit operator is diagonal in this representation, but $\hat{l}_\textrm{Yb} \cdot \hat{s}_\textrm{Rb}$ can change $m_j$ and $m_{s,\textrm{Rb}}$ by 1 in opposite directions (and can thus change $j$ and/or $f$ by 1).

\section{Effect of inelastic decay on quasibound states and Feshbach resonances}

The following sections need an understanding of how inelastic decay affects the properties of bound states and resonances, so here we give a brief summary of the key results.

In a multichannel system, true bound states occur at energies where all channels in the coupled equations are asymptotically closed, $E<E_i$ for all channels $i$, where $E_i$ is the energy of the separated particles in channel $i$. At energies above the lowest threshold $E_0$, distinct states may still exist, but they may decay to the open channels with $E_i<E$. Such states are termed \emph{quasibound}, and may be observed spectroscopically; they are characterized by their energy $E_\textrm{res}$ and width $\Gamma_E$. A quasibound state decays into the continuum with a lifetime $\tau=\hbar/\Gamma_E$.

A quasibound state appears in a scattering calculation as a Feshbach resonance. In the simplest case, with one open channel, the scattering phase shift $\delta(E)$ increases by $\pi$ across the resonance according to the Breit-Wigner formula,
\begin{equation}
\delta(E) = \delta_\textrm{bg}(E)+\arctan\left[\frac{\Gamma_E}{2(E_\textrm{res}-E)} \right],
\end{equation}
where $\delta_\textrm{bg}(E)$ is a slowly varying background (non-resonant) phase shift. Well above threshold, $\Gamma_E$ is almost independent of energy. If there is more than one open channel, similar behavior is shown by the S-matrix eigenphase sum $\mathcal{S}$, which is the sum of the phases of the complex eigenvalues of $\boldsymbol{S}(E)$ \cite{Ashton:1983}.

Ultracold collision experiments are usually carried out as a function of magnetic field, which can shift molecular states with respect to atomic thresholds. When a true bound state crosses the lowest threshold as a function of magnetic field $B$ (or any other parameter in the Hamiltonian), it causes a zero-energy Feshbach resonance. This produces a pole in the s-wave scattering length $a(B)$,
\begin{equation}
a(B) = a_\textrm{bg}(B) \left(1-\frac{\Delta}{B-B_\textrm{res}}\right),
\end{equation}
where $a_\textrm{bg}(B)$ is a slowly varying background (non-resonant) scattering length, $B_\textrm{res}$ is the resonance position and $\Delta$ is its elastic width. At sufficiently low energy, $\Delta$ is independent of energy. At the lowest threshold, $a(B)$, $a_\textrm{bg}(B)$ and $\Delta$ are all real.

The situation is considerably more complicated for scattering at a higher threshold $i$ with $E_i>E_0$, where inelastic processes are possible \cite{Hutson:res:2007}. The scattering length is then complex, $a(B) = \alpha(B) - \textrm{i} \beta(B)$ \cite{Balakrishnan:scat-len:1997}, and its imaginary part characterizes the total inelastic scattering from the incoming channel to all other channels that are energetically accessible. In the simple case where there is no background inelastic scattering, $a_\textrm{bg}(B)$ is real. However, if the quasibound state that causes the resonance can itself decay when it is below threshold, with width $\Gamma_E^\textrm{inel}$, the pole in the s-wave scattering length is replaced with an oscillation: the real part $\alpha(B)$ shows an oscillation of amplitude $\pm a_\textrm{res}/2$ and the negative imaginary part $\beta(B)$ shows a Lorentzian peak of height $a_\textrm{res}$ and width $\Gamma_B^\textrm{inel} = \Gamma_E^\textrm{inel}/\Delta\mu$; here
$\Delta\mu = d (E_i-E_\textrm{res})/dB$ is the difference in magnetic moments between the incoming threshold and the quasibound state. The resonant scattering length $a_\textrm{res} = -2 a_\textrm{bg} \Delta/\Gamma_B^\textrm{inel}$ depends on the \emph{ratio} of the coupling from the resonant state to the incoming channel and the lower-lying (inelastic) channels. Resonances of this type, due to quasibound states, are referred to as \emph{decayed} resonances;
they range from pole-like resonances where $\Gamma_B^\textrm{inel} \ll \Delta$ to resonances that are almost invisible in the scattering length where $\Gamma_B^\textrm{inel} \gg \Delta$.

The situation is further complicated where there is background inelastic scattering from the incoming channel, so that $a_\textrm{bg}(B)$ is itself complex. This includes the case of overlapping decayed resonances, where ``background" inelasticity for one resonance is provided by another nearby resonance. In this case $a_\textrm{res}$ can also be complex, and both the real and imaginary parts of the scattering length can show complicated lineshapes with both peaks and troughs. The lineshapes may nevertheless still be characterized in terms of complex $a_\textrm{bg}$ and $a_\textrm{res}$ and real $\Delta$ and $\Gamma_B^\textrm{inel}$ \cite{Frye:resonance:2017}. If $\Delta\mu$ is known, $\Gamma_B^\textrm{inel}$ may be used to extract $\Gamma_E^\textrm{inel}$ and hence the lifetime $\tau$ of the state that causes the resonance. This lifetime is important when considering the possibility of magnetoassociation at a decayed resonance, because it is the lifetime of the molecular state that would be produced.

Frye and Hutson \cite{Frye:resonance:2017} have developed methods for converging on and characterizing Feshbach resonances from coupled-channel calculations. The methods work efficiently for isolated resonances of all three cases described above: elastic resonances and decayed resonances with and without background inelasticity. However, they sometimes converge poorly in cases where resonances overlap and interfere, as happens in some of the cases here.

\section{Resonances at \protect{R\lowercase{b}}($^2$S) + \protect{Y\lowercase{b}}($^3$P$_2$) thresholds}

In zero magnetic field, an $^{87}$Rb atom has two hyperfine states, $f=1$ and 2, separated by 6.835 GHz. In a magnetic field, each of these splits into $2f+1$ sublevels, with adjacent sublevels separated by 0.7 MHz/G at low field. Yb($^3$P$_2$) similarly splits into 5 sublevels, with adjacent levels separated by 2.1 MHz/G at low field.

Any collision between Rb($^2$S) + Yb($^3$P$_2$) can result in inelastic transitions. Even for the lowest such threshold, with $(f,m_f)=(1,1)$ and $m_j=-2$, inelastic collisions to Yb($^3$P$_0$) can be driven by $V_2^+(R)$ and $V_2^-(R)$, while inelastic collisions to Yb($^3$P$_1$) can be driven by $V_0^-(R)$, $V_2^+(R)$, $V_2^-(R)$ and $\Delta a_\textrm{Rb}(R)$. For higher thresholds, there are additional inelastic mechanisms that change $f$, $m_f$ or $m_j$.

For alkali-metal pairs, inelastic collisions are generally suppressed for atom pairs initially in spin-stretched and quasi-spin-stretched channels, even when they are energetically allowed. Spin-stretched channels are those that have the maximum values of $m_f$ for both atoms, while quasi-spin-stretched channels are those for which there is no lower channel with the same value of $M_F=m_f+m_{f,\textrm{Yb}}$; for bosonic Yb isotopes with nuclear spin 0, $m_{f,\textrm{Yb}}=m_j$. Inelastic processes from such channels are suppressed for two reasons. First, collisions that change $M_F$ are driven only by anisotropic terms in the interaction potential, which are weak for alkali-metal pairs. Secondly, for incoming s-wave collisions, an outgoing wave for a different value of $M_F$ must have $L\ge 2$; this creates a centrifugal barrier in the outgoing channel that is often sufficient to cause further suppression of the inelastic rate when the energy release is small.

\begin{figure}[tbp]
	\subfloat[]{
		\includegraphics[width=0.45\textwidth]{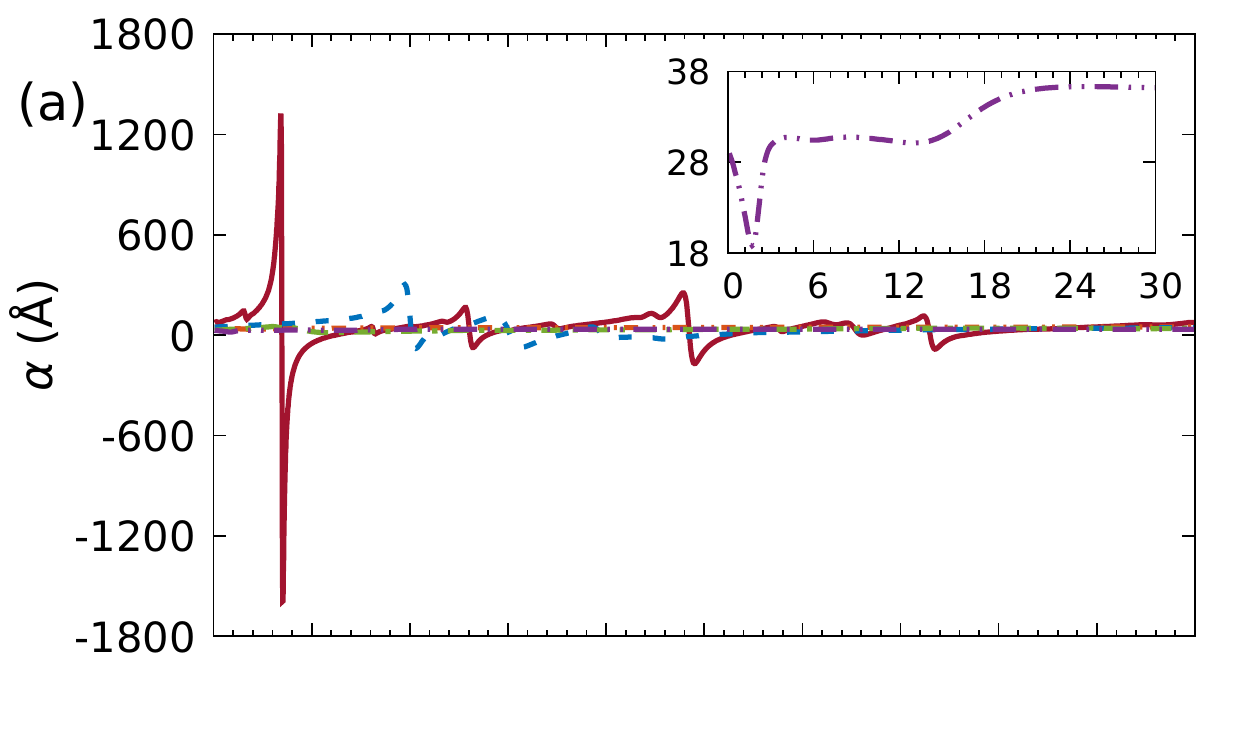}
	}
    %%%%%%%%%%%%%%%%%%%%%%%%%%%%%%%%%%%%second row
    \vspace{-1 cm}
	\subfloat[]{
		\includegraphics[width=0.45\textwidth]{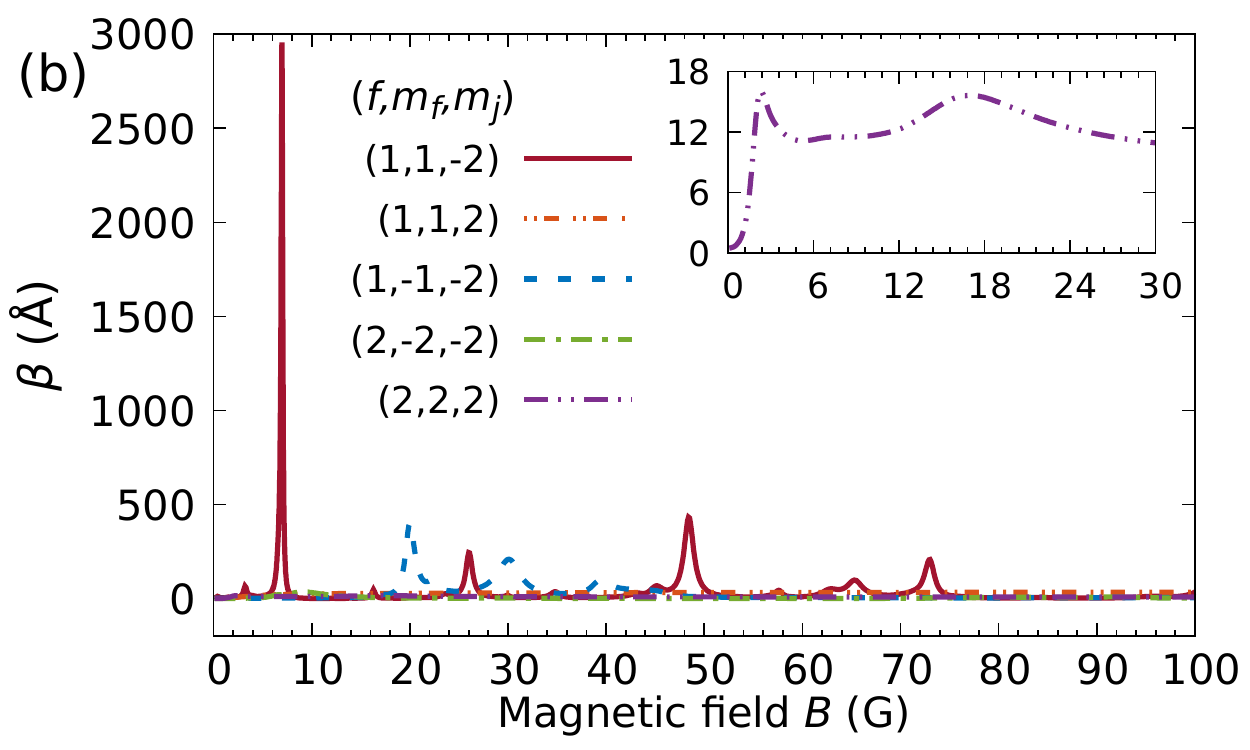}
	}
    \vspace{-0.5 cm}
\caption{Real (a) and negative imaginary (b) parts of the scattering length for Rb$(f,m_f)$ colliding with $^{170}$Yb($^3$P$_2,m_j$) for different initial atomic states. The insets shows the weak features in the spin-stretched channel at low field.}%
\label{fig:channels3P2}
\end{figure}

Spin-stretched channels behave quite differently for Rb($^2$S) + Yb($^3$P$_2$). $V_2^+$ provides a strong anisotropic interaction that can change $m_j$ and $m_L$ by 1 or 2 in opposite directions, while conserving their sum. Although it can change $j$, it also has matrix elements diagonal in $j$ for $j=1$ and 2. Such a term does \emph{not} conserve $m_f+m_j$. $V_2^-$ has even weaker selection rules, and can change $f$ and/or $m_f$ by 1 at the same time, conserving only $M_\textrm{tot}$. For incoming s-wave collisions, there can still be some centrifugal suppression at low fields due to the centrifugal barrier in the outgoing channel, which for $L=2$ is approximately $h\times6$~MHz high. At higher fields, inelastic losses cause substantial damping of resonant poles, even for quasi-spin-stretched states.

We have carried out coupled-channel scattering calculations on Rb+$^{170}$Yb($^3$P$_2$), using the \textsc{molscat} package \cite{molscat:2019, mbf-github:2020}, for a variety of incoming channels labeled by $f$, $m_f$ and $m_j$. The real and imaginary parts of the scattering length are shown as a function of $B$ in Fig.\ \ref{fig:channels3P2}. There are many resonances, but most of them are strongly decayed. In general, resonances with large values of $a_\textrm{res}$ [and thus high peaks in $\beta(B)$] are likely to be the most promising for magnetoassociation. Only the lowest $^3$P$_2$ channel, with $(f,m_f,m_j)=(1,1,-2)$, shows large-amplitude features in the scattering length, with $a_\textrm{res} \approx 3000$~\AA\
here. The spin-stretched channel (2,2,2) is the highest channel for Yb($^3$P$_2$), so shows no resonant features at all, except below $\sim20$~G, where states with $L>0$ confined behind their centrifugal barrier can contribute. Even the quasi-spin-stretched channel $(1,-1,-2)$ shows only weak resonant features with $a_\textrm{res} < 400$~\AA.

\begin{figure}[tbp]
	\subfloat[]{
		\includegraphics[width=0.5\textwidth]{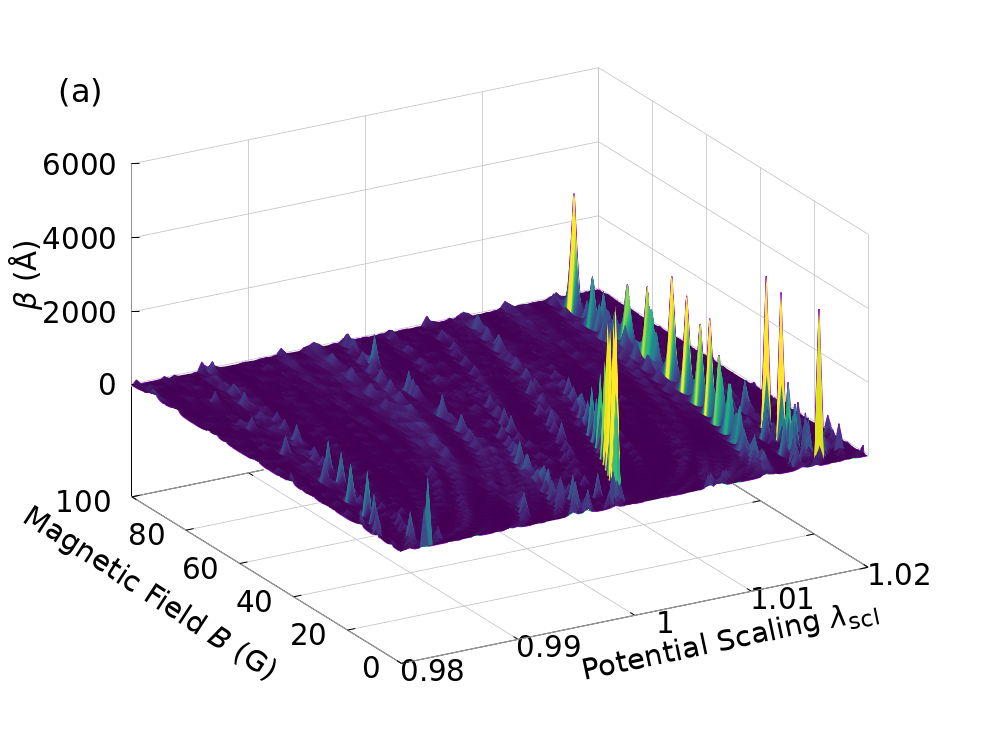}
	}
    %%%%%%%%%%%%%%%%%%%%%%%%%%%%%%%%%%%%second row
    \vspace{-0.5 cm}
	\subfloat[]{
		\includegraphics[width=0.45\textwidth]{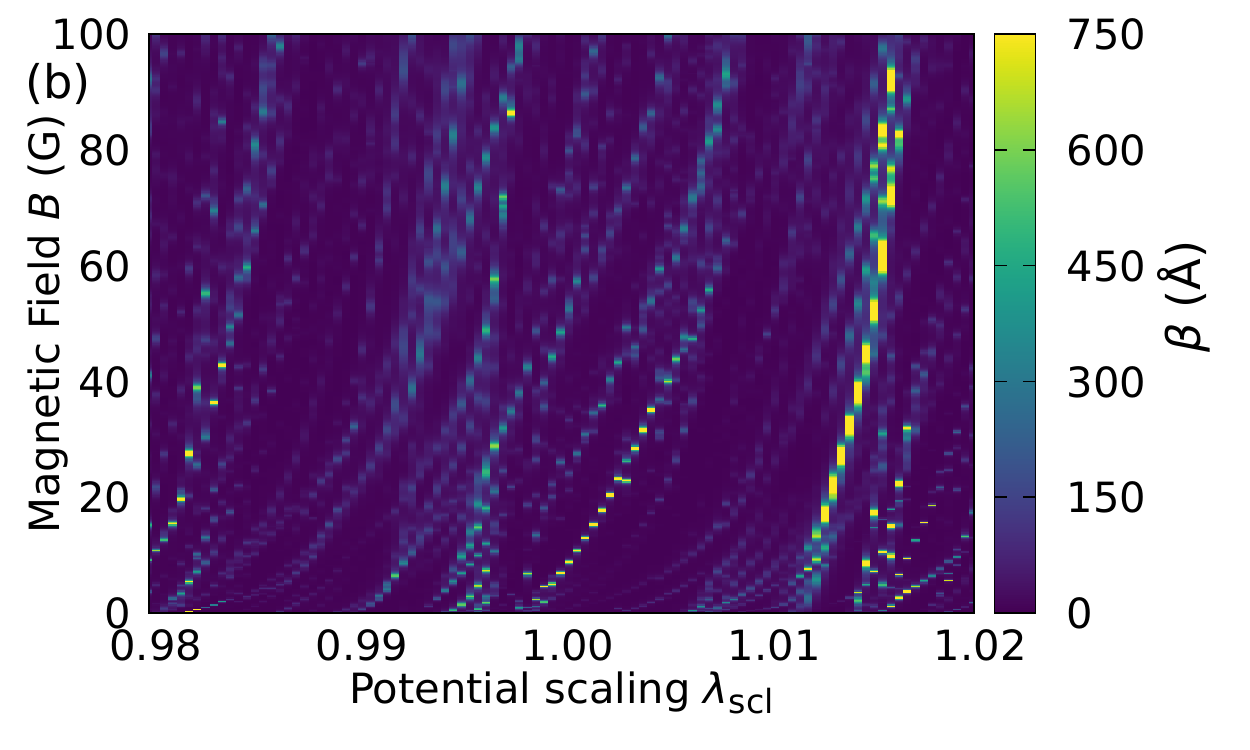}
	}
    \vspace{-0.5 cm}
\caption{(a) Negative imaginary part of scattering length, $\beta$, for Rb($f=1,m_f=1$) colliding with $^{170}$Yb($^3$P$_2,m_j=-2$), as a function of magnetic field and potential scaling; (b) same results shown as a contour plot. The sawtooth appearance of the ridges as a function of field is an artifact of the plotting procedure.
}%
\label{fig:lambda-scan-170}
\end{figure}

Our calculations so far have been on a single interaction potential (and set of spin-orbit functions) fitted to the electronic structure calculations of Shundalau and Minko \cite{Shundalau:2017}. However, any such interaction potential has intrinsic uncertainties, which in this case are many percent. It is therefore important to investigate the sensitivity of the results to the interaction potential. This is a many-dimensional space, but to sample it in a systematic way we investigate simple scalings of the entire interaction operator by a factor $\lambda_\textrm{scl}$, which we vary over a range of 2\% around the original. Figure \ref{fig:lambda-scan-170} shows the imaginary part of the scattering length for $^{87}$Rb($^2$S,$f=1,m_f=1)$ + $^{170}$Yb($^3$P$_2$) for 100 different values of $\lambda_\textrm{scl}$ across this range. It may be seen that different scalings of the potential give quite different patterns of resonances. Nevertheless, the positions, widths and amplitudes of the resonances are reasonably smooth functions of $\lambda_\textrm{scl}$. Without detailed experiments to refine the interaction potentials, there is no way of knowing where on this plot the real system will fall. Indeed, since Fig.\ \ref{fig:lambda-scan-170} shows only a 1-dimensional cut through a many-dimensional space of potential parameters, it is unlikely that any single value of $\lambda_\textrm{scl}$ will reproduce the real behavior. Nevertheless, the cuts shown give a reasonable sample of the likely behaviors.

\begin{figure}[tbp]
	\subfloat[]{
		\includegraphics[width=0.5\textwidth]{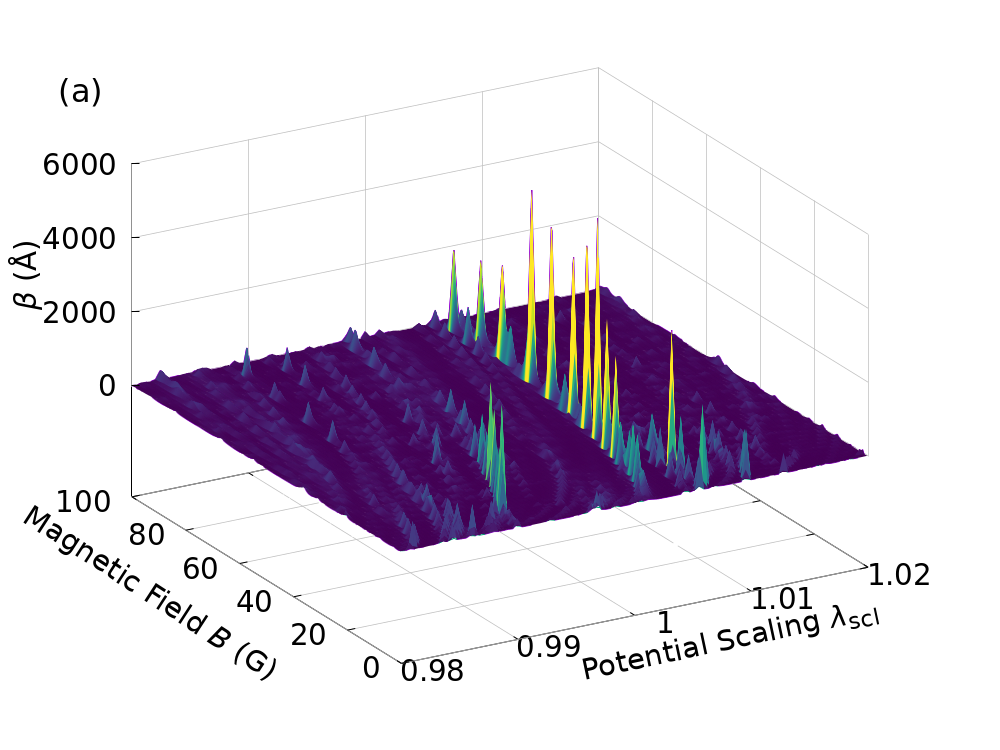}
	}
    %%%%%%%%%%%%%%%%%%%%%%%%%%%%%%%%%%%%second row
    \vspace{-0.5 cm}
	\subfloat[]{
		\includegraphics[width=0.45\textwidth]{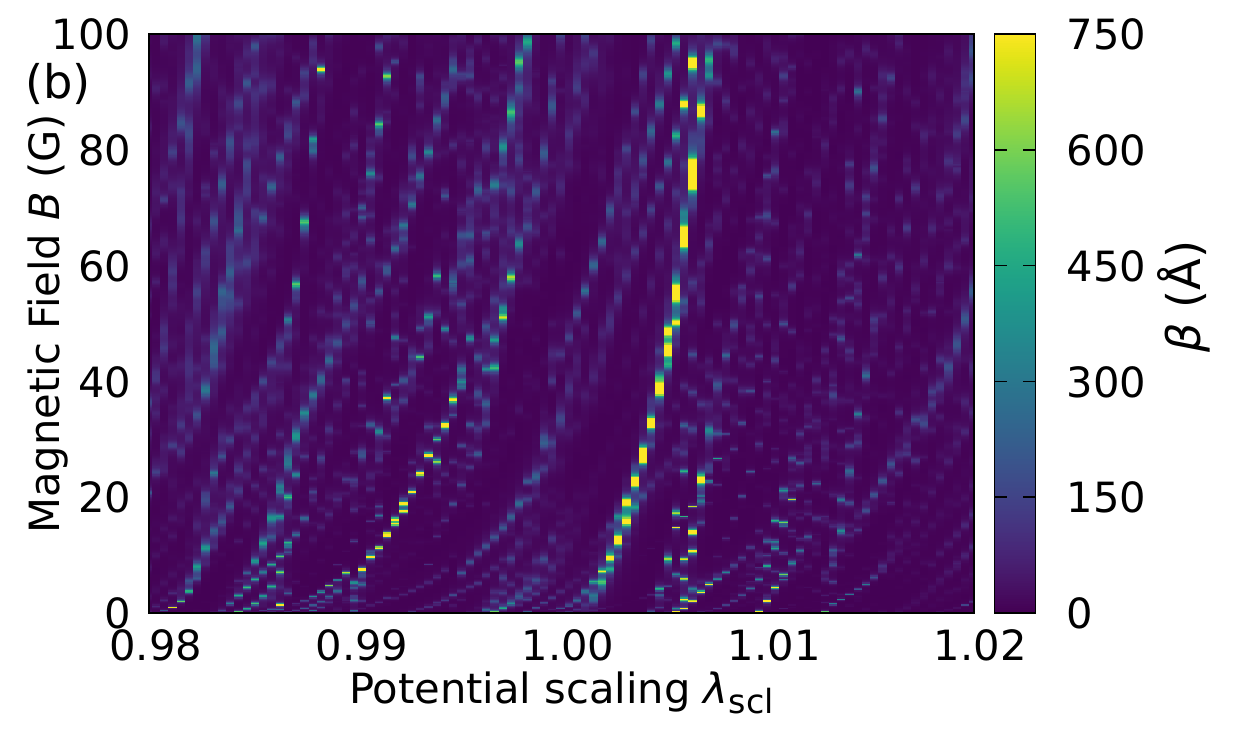}
	}
    \vspace{-0.5 cm}
\caption{(a) Negative imaginary part of scattering length, $\beta$, for Rb($f=1,m_f=1$) colliding with $^{174}$Yb($^3$P$_2,m_j=-2$), as a function of magnetic field and potential scaling;(b) same results shown as a contour plot.}%
\label{fig:lambda-scan-174}
\end{figure}

The coupled equations \eqref{eq:se-invlen} contain matrix elements of the mass-scaled interaction operator $2\mu\hat{V}/\hbar^2$. For this reason, scaling the interaction operator $\hat{V}$ by a factor $\lambda_\textrm{scl}$ has almost the same effect as scaling the reduced mass $\mu$ by the same factor.  The only difference arises because $\lambda_\textrm{scl}$ does not scale $\hat{H}_\textrm{Rb}$ and $\hat{H}_\textrm{Yb}$ in the same way; this has a minor effect on inelastic processes, but very little effect on elastic processes. Figure \ref{fig:lambda-scan-174} shows the same plot as Fig.\ \ref{fig:lambda-scan-170}, but for $^{174}$Yb, which has a reduced mass 0.78\% larger than $^{170}$Yb. The plots confirm that the pattern of resonances is very similar, but shifted down in scaling factor by the expected 0.78\%.

\begin{figure}[tbp]
	\subfloat[]{
		\includegraphics[width=0.45\textwidth]{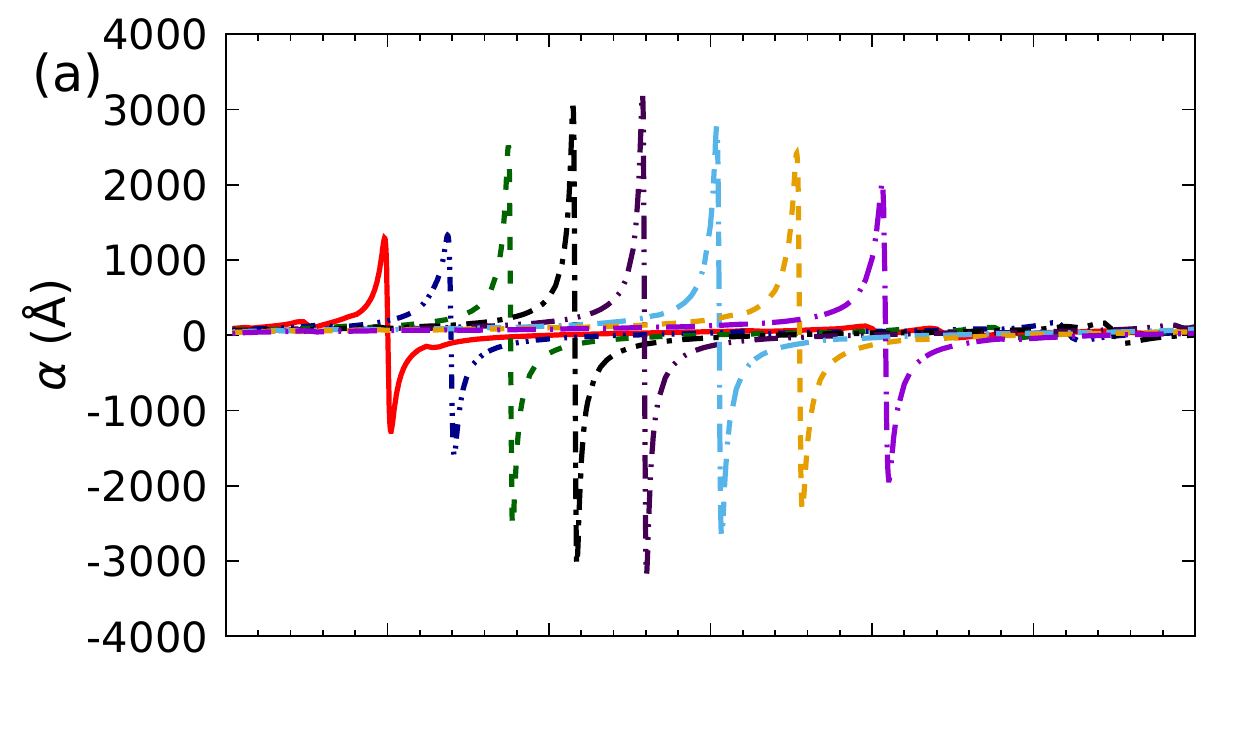}
	}
    %%%%%%%%%%%%%%%%%%%%%%%%%%%%%%%%%%%%second row
    \vspace{-1 cm}
	\subfloat[]{
		\includegraphics[width=0.45\textwidth]{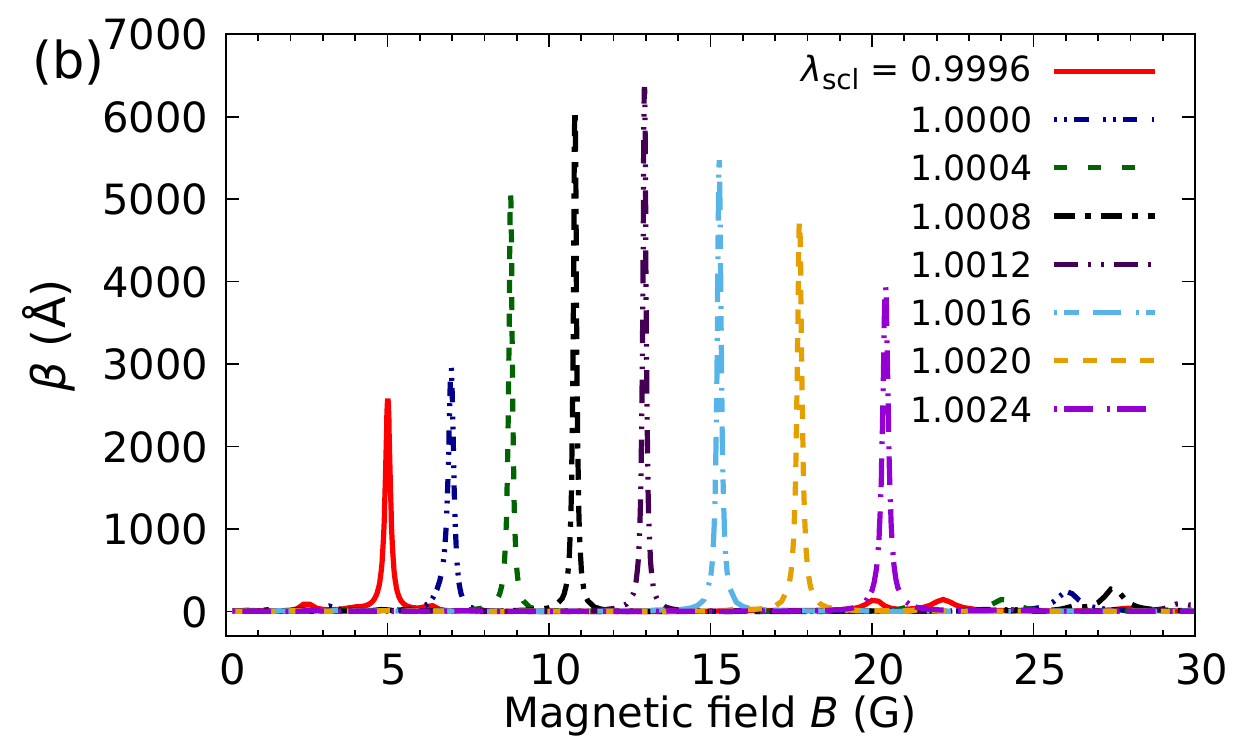}
	}
    \vspace{-0.5 cm}
\caption{Resonances in the real (a) and negative imaginary (b) parts of the scattering length for Rb$(f=1,m_f=1)$ colliding with $^{170}$Yb($^3$P$_2,m_j=-2$) for different scalings of the interaction potential.}%
\label{fig:a-scan-3P2}
\end{figure}

We thus see that about 30\% of potentials show fairly sharp resonances in the scattering length for the lowest threshold of Rb($^2$S) + Yb($^3$P$_2$) at fields below 100~G. This occurs for any one isotope of Yb, and there are 5 bosonic and 2 fermionic isotopes available, so there is a good likelihood that one or more isotopes will show such resonances for the real interaction potential.

We now turn to the question of whether such resonances are likely to be useful for molecule formation.
To assess this, we pick potentials with $\lambda_\textrm{scl}$ from 0.9996 to 1.0024. These are not the highest features in Fig.\ \ref{fig:lambda-scan-170}, but they exist over a relatively broad range of interaction potentials. The resonant features in the scattering length are shown in Fig.\ \ref{fig:a-scan-3P2}, where it may be seen that the peaks in $\beta(B)$ are up to 6000 \AA\ high.

For each value of $\lambda_\textrm{scl}$, we characterize the sharpest resonance, using the regularized scattering length procedure of Frye and Hutson \cite{Frye:resonance:2017}, as implemented in \textsc{molscat} \cite{molscat:2019, mbf-github:2020}. The parameters we obtain are the resonance position $B_\textrm{res}$, the background scattering length $a_\textrm{bg}$, the elastic width $\Delta$, the resonant scattering length $a_\textrm{res}$ and the inelastic resonance width $\Gamma_B^\textrm{inel}$. These parameters are tabulated in Supplemental Material \cite{sup-mat-RbYb}. They show some irregularity because of the presence of overlapping resonances that are different for each scaling, but the key parameter for the present purpose is $\Gamma_B^\textrm{inel}$, which is stable. The smallest value calculated for $\Gamma_B^\textrm{inel}$ is about 0.125~G. This is related to the energy width of the molecular state produced by magnetoassociation, $\Gamma_E^\textrm{inel} = \Gamma_B^\textrm{inel}\,\Delta\mu$, where $\Delta\mu$ is the gradient at which the quasibound state responsible for the resonance crosses threshold.

We have also characterized the resonant state just below threshold. For the case of $\lambda_\textrm{scl}=1.0012$, the state crosses threshold and causes a zero-energy Feshbach resonance near $B=13$~G. At fields just below this, there are clear Breit-Wigner signatures in the calculated eigenphase sum below threshold. We have characterized these as described in Supplemental Material \cite{sup-mat-RbYb} between 9~G and 10~G, using the method of Frye and Hutson \cite{Frye:quasibound:2020}. We obtain an energy gradient $\Delta\mu \approx -0.5$ MHz/G and energy widths close to threshold $\Gamma_E^\textrm{inel}\approx$ 0.06 MHz. These values are consistent with the observed values of $\Gamma_B^\textrm{inel}$ described above. We indicate quantum numbers of the state that crosses threshold with subscript res; the low value of $\Delta\mu$ suggests that the resonant state here has predominantly $(m_{f,\textrm{res}},m_{j,\textrm{res}})=(0,-2)$ and thus $L_\textrm{res}\ge2$.

An energy width $\Gamma_E^\textrm{inel} \approx 0.06$ MHz corresponds to a lifetime of the molecular state $\tau = \hbar/\Gamma_E^\textrm{inel} \approx 2.5~\mu$s. In view of the uncertainty in the interaction potential, this should not be considered more than an order-of-magnitude estimate. Nevertheless it is short enough that it is likely to be experimentally challenging to transfer molecules made by magnetoassociation at the $^3$P$_2$ thresholds to a stabler state before they undergo inelastic decay. However, even if they do not prove useful for molecule formation, resonances at these thresholds may be used to tune the scattering length by substantial amounts, and may be useful in future experiments.

As mentioned above, there are two other resonances that produce prominent ridges in Figs.\ \ref{fig:lambda-scan-170} and \ref{fig:lambda-scan-174}. One of these has higher peaks in $\beta(B)$, with smaller values of $\Gamma_B^\textrm{inel}$ and correspondingly longer molecular lifetimes. However, it exists for only a tiny range of scaling factors and magnetic fields, so is unlikely to be seen in the real system. The other exists over a wider range of scaling factor, but is more strongly decayed, with larger $\Gamma_B^\textrm{inel}$ and shorter lifetimes. Both these are described in the Supplemental Material \cite{sup-mat-RbYb}.

\section{Resonances at {R\lowercase{b}}($^2$S) + {Y\lowercase{b}}($^3$P$_0$) thresholds}

A Yb atom in its $^3$P$_0$ state is spherical. Its collisions with Rb($^2$S) have many similarities to those of Yb($^1$S). The scattering is governed mostly by a single effective potential curve, the lowest in Fig.\ \ref{fig:pot-curves}(a). The resulting scattering length varies only slowly with magnetic field $B$, except near narrow Feshbach resonances. Each near-threshold bound state is almost parallel to the Rb hyperfine state that supports it as a function of $B$. All the mechanisms that can produce Feshbach resonances for Rb+Yb($^1$S) \cite{Zuchowski:RbSr:2010, Brue:LiYb:2012, Brue:AlkYb:2013, Barbe:RbSr:2018, Yang:CsYb:2019} exist for Yb($^3$P$_0$) as well, with coupling between the bound states and thresholds provided by the dependence of the Rb or Yb hyperfine coupling on internuclear distance $R$. However, there are additional mechanisms for Rb+Yb($^3$P$_0$) due to the additional terms in the interaction potential.

The $^3$P$_1$ and $^3$P$_2$ states of Yb lie 703.568 and 2421.949 cm$^{-1}$, respectively, above $^3$P$_0$. Any bound states that they support will be widely separated at the energy of $^3$P$_0$, and very unlikely to cause zero-energy Feshbach resonances at experimentally accessible magnetic fields. However, as seen above, the anisotropic and spin-dependent parts of the interaction potential are substantial compared to the spin-orbit splittings, so they cause significant mixing of the $^3$P$_0$, $^3$P$_1$ and $^3$P$_2$ states at short range. They also cause mixing between the $f=1$ and 2 hyperfine states of Rb. Because of this, zero-energy Feshbach resonances can exist where bound states supported by both $f=1$ and $f=2$ states of Rb, with both $L=0$ and $L=2$, cross thresholds with different ($f,m_f$).

\begin{figure}[tbp]
\centering
\includegraphics[width=0.5\textwidth]{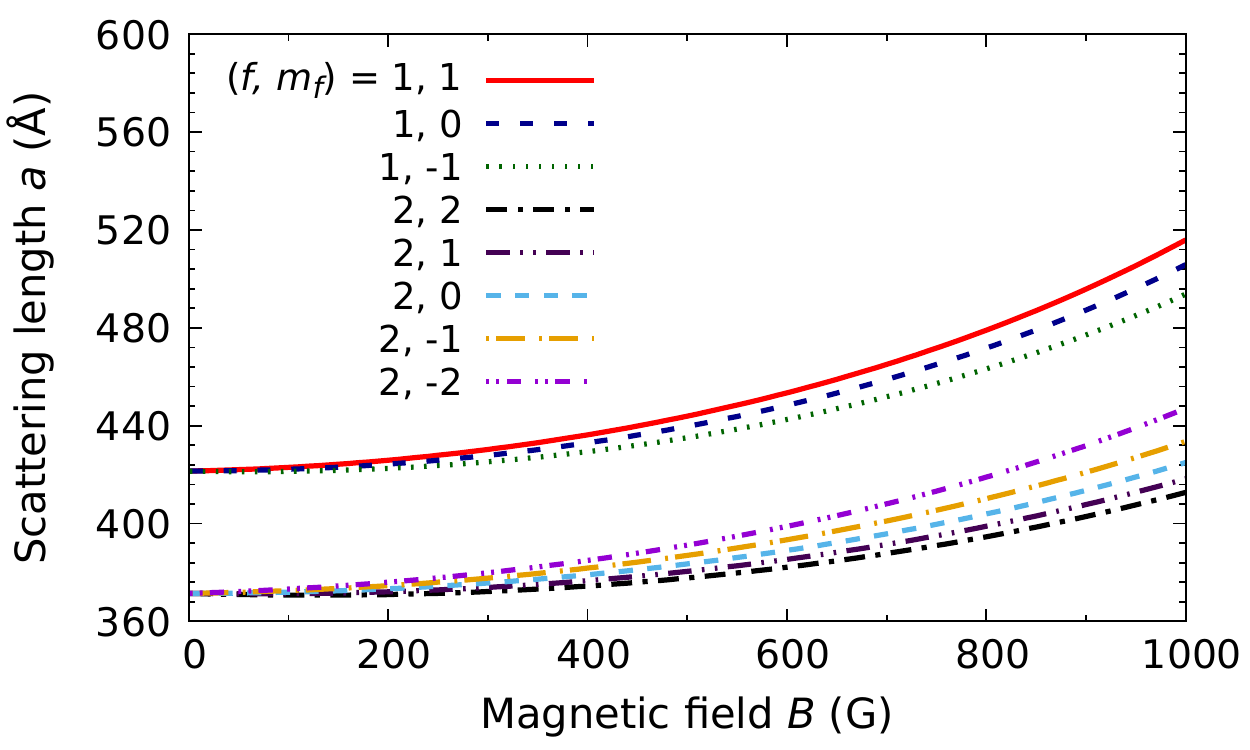}%
\caption{Real part of the scattering length for different hyperfine states of Rb$(f,m_f)$ colliding with $^{174}$Yb($^3$P$_0$).}%
\label{fig:a-scan-3P0}
\end{figure}

We have calculated scattering lengths for a variety of thresholds for Rb($f,m_f$) interacting with $^{174}$Yb($^3$P$_0$), using the same methods as for Yb($^3$P$_2$) above. The results are shown in Fig.\ \ref{fig:a-scan-3P0}, initially on a coarse scale that is not designed to show narrow Feshbach resonances. It may be seen that the scattering lengths vary only slowly with field, as expected, but that there are significant differences in the values at the different thresholds. These arise because the matrix elements involving $V_0^-$, $V_2^-$ and $\Delta a_\textrm{Rb}$ that connect the $j=0$ thresholds to $j=1$ and 2 are different for different values of $(f,m_f)$.

\begin{figure}[tbp]
\centering
\includegraphics[width=0.5\textwidth]{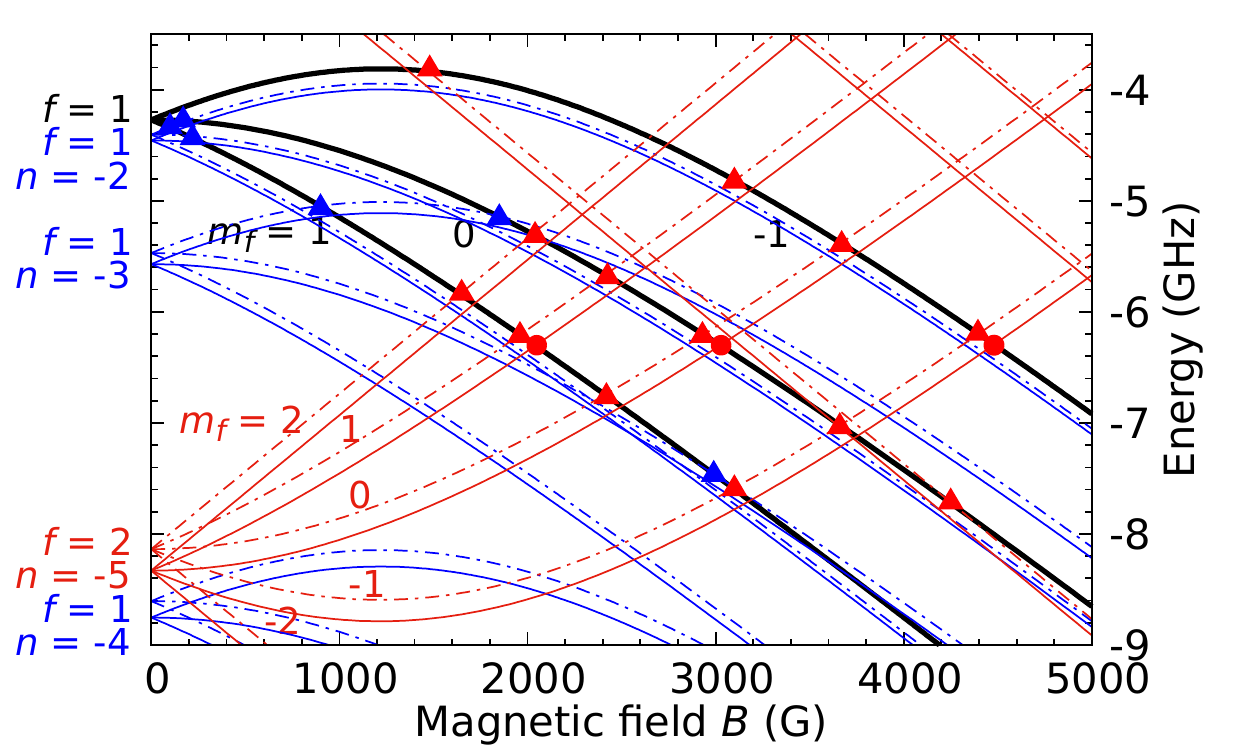}%
\caption{Level crossing diagram for Rb + $^{174}$Yb with the unscaled interaction potential. The heavy black lines show the hyperfine thresholds for $f = 1$. The quantum numbers $f$ and $n$ are given on the left-hand side for each manifold of molecular levels (thin colored lines, solid for $L = 0$ and dashed for $L = 2$). Crossings that cause Feshbach resonances are indicated with symbols as described in the text.
}%
\label{fig:crossing}
\end{figure}

The scattering lengths allow us to calculate zero-field binding energies for states below the $f=1$ and $f=2$ thresholds. For this we use single-channel calculations on the lowest adiabat shown in Fig.\ \ref{fig:pot-curves}(a), with small adjustments at short range to match the required scattering length. We then use the zero-field binding energies to calculate the pattern of states below each threshold, with the initial approximation that each state is parallel to the threshold that supports it. This approximation could be improved, using the calculated $B$-dependence of the scattering lengths, but it is conceptually useful. Figure \ref{fig:crossing} shows the resulting diagram for $^{174}$Yb, including the $f=1$ thresholds themselves.

Many crossings between bound states and thresholds are visible in Fig.\ \ref{fig:crossing}. To understand which of these can cause Feshbach resonances, we must consider the couplings due to different terms in the interaction potential. First, $V_0^-$ can couple $j=0$ to $j=1$ differently for $f=1$ and 2. However, it cannot change $m_f+m_j$, so even in second order it cannot change $m_f$ at the $j=0$ threshold, where $m_j$ is always zero. A similar argument applies to $\Delta a_\textrm{Rb}$. Bound states with $f_\textrm{res}=1$ are parallel to thresholds with $f=1$ and the same $m_f$, so do not cross them. However, bound states with $f_\textrm{res}=2$ can cross thresholds with $f=1$ and the same $m_f$. Figure \ref{fig:crossing} shows one crossing of this type at each threshold as a red circle, and zero-energy Feshbach resonances are expected at these fields. These are due to a state bound by 10.9 GHz with respect to the $f=2$ threshold, with vibrational quantum number $n=-5$ relative to that threshold. We refer to such resonances as $L$-conserving and $f$-changing.

The spin-averaged anisotropic term $V_2^+$ can couple $j=0$ to $j=2$, but the couplings are independent of $f$ and $m_f$. It therefore does not cause zero-energy Feshbach resonances in second order, though it can do so in higher order in combination with other terms such as $V_0^-$. However, the spin-difference anisotropic term $V_2^-$ has couplings that are different for $f=1$ and 2. Moreover, it can change $L$ and $M_L$ in addition to $f$ and $m_f$. In second order, it can couple a threshold $(f,m_f$) to a bound state with $f_\textrm{res}=1$ or 2 and $L_\textrm{res}=0$ or 2. For a bound state with $L_\textrm{res}=2$, $m_{f,\textrm{res}}$ can take values $m_f, m_f\pm1, m_f\pm2$, compensated by $M_{L,\textrm{res}}=0,\mp1,\mp2$. This allows resonances due to bound states with both $f=2$ and $f=1$ at thresholds with $f=1$, in a similar way to resonances arising from mechanism III at the $^1$S thresholds in fermionic isotopes of RbSr \cite{Barbe:RbSr:2018}, CsYb \cite{Yang:CsYb:2019} and LiYb \cite{Green:LiYb-res:2020}. We refer to such resonances as $L$-changing and either $f$-conserving or $f$-changing; the corresponding crossings are identified in Fig.\ \ref{fig:crossing} with blue and red triangles, respectively. Since there are always bound states with $f_\textrm{res}=1$ and $L_\textrm{res}=2$ quite close to the $f=1$ threshold, $f$-conserving but $L$-changing Feshbach resonances will always exist at experimentally accessible fields, even for bosonic isotopes of Yb.

We have located all the zero-energy Feshbach resonances shown in Fig.\ \ref{fig:crossing} at the threshold $(f,m_f)=(1,1)$ by performing coupled-channel bound-state calculations as a function of magnetic field at zero energy using the \textsc{field} package \cite{bound+field:2019, mbf-github:2020}. We have then characterized the resonances using the elastic procedure of Frye and Hutson \cite{Frye:resonance:2017}, as implemented in \textsc{molscat} \cite{molscat:2019, mbf-github:2020}. The resulting resonant fields and widths are given in Table \ref{tab:res-3P0}. It may be seen that the $f$-changing but $L$-conserving resonance has a calculated width of 76 mG, and even one of the $L$-changing resonances is 10~mG wide.

\begin{table}[tbp]
\caption{Calculated parameters for resonances at the threshold Rb($f=1,m_f=1$) + $^{174}$Yb($^3$P$_0$) on the unscaled interaction potential.}
\begin{footnotesize}
\begin{ruledtabular}
\begin{tabular}{cccc}
$L_{\rm res},f_{\rm res},m_{f,{\rm res}}$ & $B_{\rm res}$ (G) & $\Delta$ (G) & $a_{\rm bg}$ ({\AA}) \\
\hline
$2, 1, -1$ & 95 & $1.2 \times 10^{-7}$ & 382 \\
$2, 1, 0$ & 193 & $3.1 \times 10^{-6}$ & 429 \\
$2, 1, -1$ & 898 & $6.7 \times 10^{-7}$ & 519 \\
$2, 2, 2$ & 1667 & $1.0 \times 10^{-2}$ & 828 \\
$2, 2, 1$ & 1986 & $6.1 \times 10^{-3}$ & 1380 \\
$0, 2, 1$ & 2070 & $7.6 \times 10^{-2}$ & 1720 \\
$2, 2, 0$ & 2438 & $-4.4 \times 10^{-3}$ & $-9100$ \\
$2, 1, 0$ & 2977 & $-7.2 \times 10^{-3}$ &  $-733$ \\
$2, 2, -1$ & 3119 & $-2.0 \times 10^{-7}$ & $-995$ \\
\end{tabular}
\end{ruledtabular}
\label{tab:res-3P0}
\end{footnotesize}
\end{table}

There are additional resonances at higher thresholds of Rb, as shown in Fig.\ \ref{fig:crossing}. We have located the resonances at the thresholds $(f,m_f)=(1,0)$ and $(1,-1)$ and characterized them using the weakly inelastic procedure \cite{Frye:resonance:2017}, which is appropriate when the background inelasticity is weak. The results are given in Tables \ref{tab:res-3P0_f1mf0} and \ref{tab:res-3P0_f1mf-1}. Some of the resonances are strongly decayed, but others have very small values of $\Gamma_B^\textrm{inel}$ and may be suitable for molecule formation. The $L$-conserving resonance at the $(1,0)$ threshold is particularly notable, with a width over 100~mG.

\begin{table*}[htbp]
\caption{Calculated parameters for resonances at the threshold Rb($f=1,m_f=0$) + $^{174}$Yb($^3$P$_0$) on the unscaled interaction potential.
}
\begin{footnotesize}
\begin{ruledtabular}
\begin{tabular}{cccccc}
$L_{\rm res},f_{\rm res},m_{f,{\rm res}}$ & $B_{\rm res}$ (G) & $\Delta$ (G) & $a_{\rm bg}$ ({\AA})
& $a_{\rm res}$ ({\AA}) & $\Gamma_B^{\rm inel}$ (G) \\
\hline
$2, 1, -1$ & 186 & $1.0 \times 10^{-6}$ & 438   & 24449 & $-3.60 \times 10^{-8}$ \\
$2, 1, -1$ & 1840 & $2.9 \times 10^{-3}$ & 981   & $1.16 \times 10^{7}$ & $-4.84 \times 10^{-7}$ \\
$2, 2, 2$ & 2059 & $5.3 \times 10^{-5}$ & 1480   & 30.9 & $-5.12 \times 10^{-3}$ \\
$2, 2, 1$ & 2447 & $-0.562$             & $-18100$ & 5.61 $\times$ 10$^{5}$ & $-3.63 \times 10^{-2}$ \\
$2, 2, 0$ & 2965 & $-1.2 \times 10^{-2}$ & $-801$ & 18400 & $-1.06 \times 10^{-3}$ \\
$0, 2, 0$ & 3056 & $-0.110$              & $-674$ & $4.98 \times 10^{6}$ & $-2.98 \times 10^{-5}$ \\
$2, 2, -1$ & 3689 & $-3.5 \times 10^{-4}$ & $-294$ & 1.49 $\times$ 10$^{6}$ & $-1.36 \times 10^{-7}$ \\
$2, 2, -2$ & 4322 & $6.1 \times 10^{-6}$ & $-119$ & 7.70 $\times$ 10$^{5}$ & $1.90 \times 10^{-9}$ \\
\end{tabular}
\end{ruledtabular}
\label{tab:res-3P0_f1mf0}
\end{footnotesize}
\end{table*}

\begin{table*}[htbp]
\caption{Calculated parameters for resonances at the threshold Rb($f=1,m_f=-1$) + $^{174}$Yb($^3$P$_0$) on the unscaled interaction potential.}
\begin{footnotesize}
\begin{ruledtabular}
\begin{tabular}{cccccc}
$L_{\rm res},f_{\rm res},m_{f,{\rm res}}$ & $B_{\rm res}$ (G) & $\Delta$ (G) & $a_{\rm bg}$ ({\AA})
& $a_{\rm res}$ ({\AA}) & $\Gamma_B^{\rm inel}$ (G) \\
\hline
$2, 2, -2$ & 1463 & $-3.1 \times 10^{-3}$ & $629$ & $3.98 \times 10^{5}$ & $9.92 \times 10^{-6}$ \\
$2, 2, 1$ & 3139 & $-3.8 \times 10^{-5}$ & $-628$ & 2.09 & $-2.24 \times 10^{-2}$ \\
$2, 2, 0$ & 3700 & $-2.4 \times 10^{-2}$ & $-302$ & 118 & $-0.122$ \\
$2, 2, -1$ & 4427 & $-4.8 \times 10^{-3}$  & $-162$ & 4440 & $-3.53 \times 10^{-4}$ \\
$0, 2, -1$ & 4511 & $-5.3 \times 10^{-2}$ & $-153$ & 2.78 $\times$ 10$^{5}$ & $-5.85 \times 10^{-5}$ \\
\end{tabular}
\end{ruledtabular}
\label{tab:res-3P0_f1mf-1}
\end{footnotesize}
\end{table*}

\begin{figure}[tbp]
	\subfloat[]{
		\includegraphics[width=0.45\textwidth]{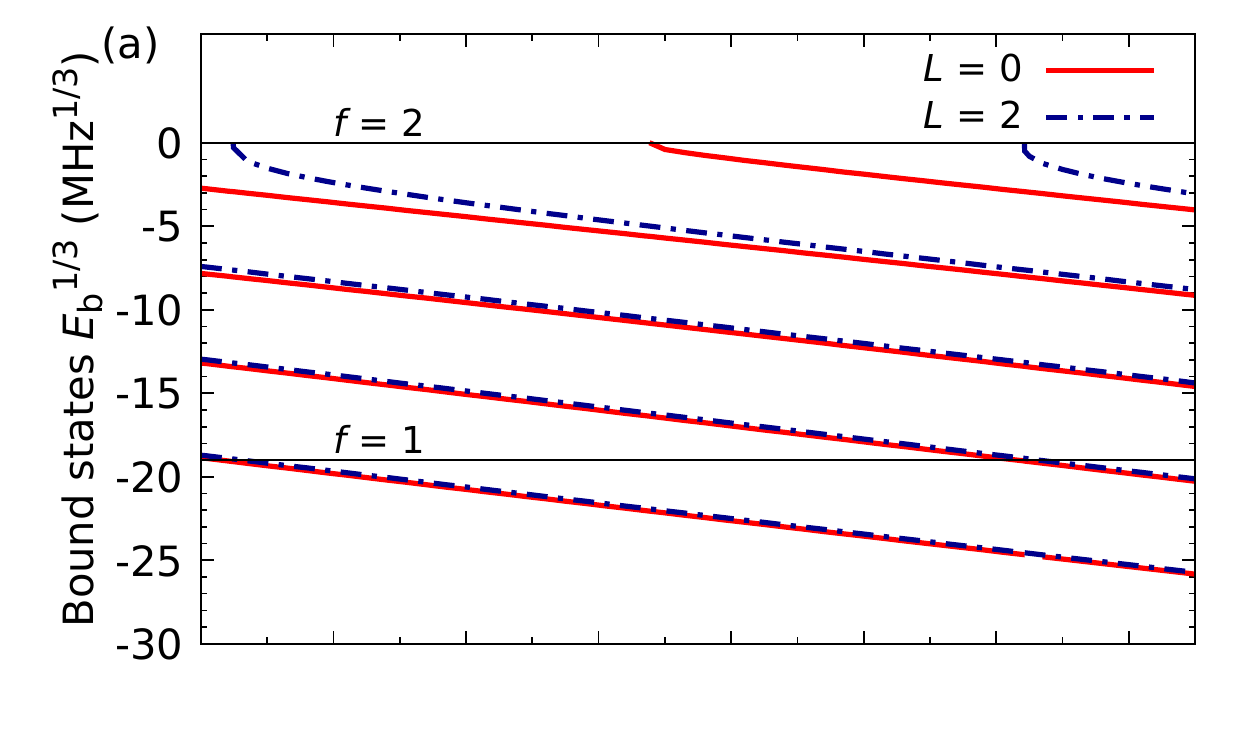}
	}
    %%%%%%%%%%%%%%%%%%%%%%%%%%%%%%%%%%%%second row
    \vspace{-1 cm}
	\subfloat[]{
		\includegraphics[width=0.45\textwidth]{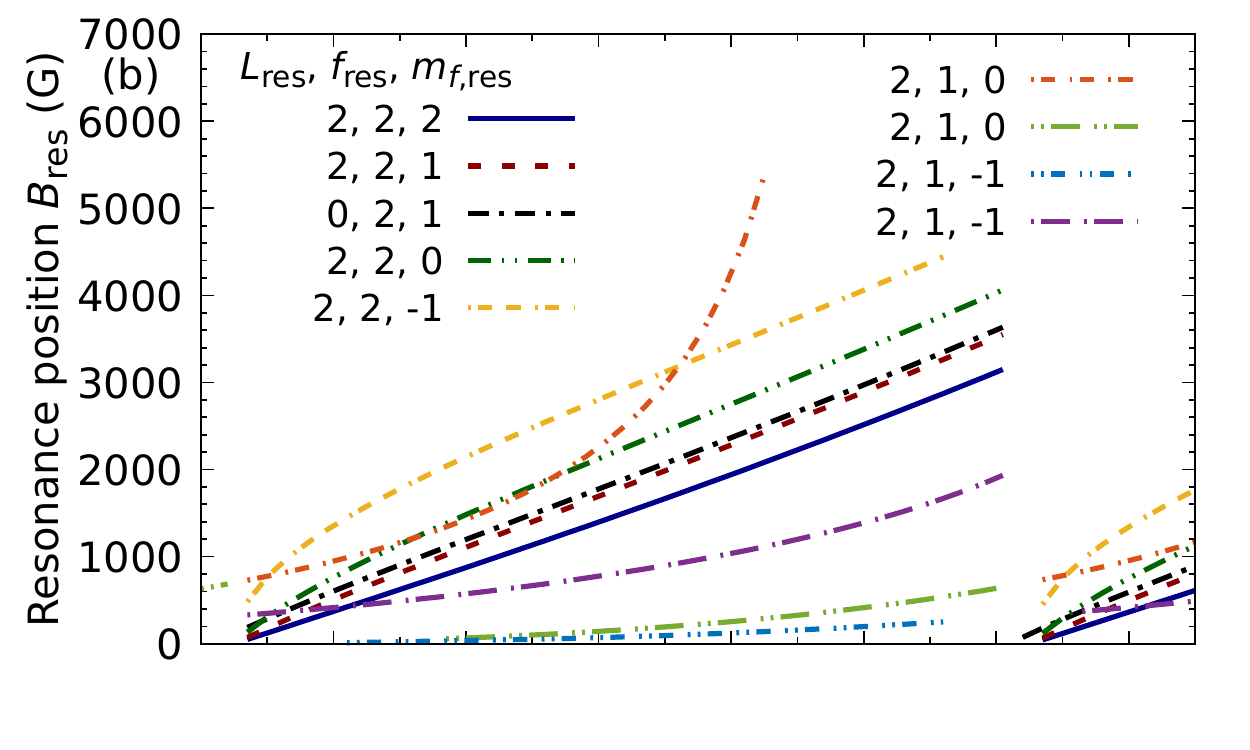}
	}
    %%%%%%%%%%%%%%%%%%%%%%%%%%%%%%%%%%%%third row
    \vspace{-1 cm}
	\subfloat[]{
		\includegraphics[width=0.45\textwidth]{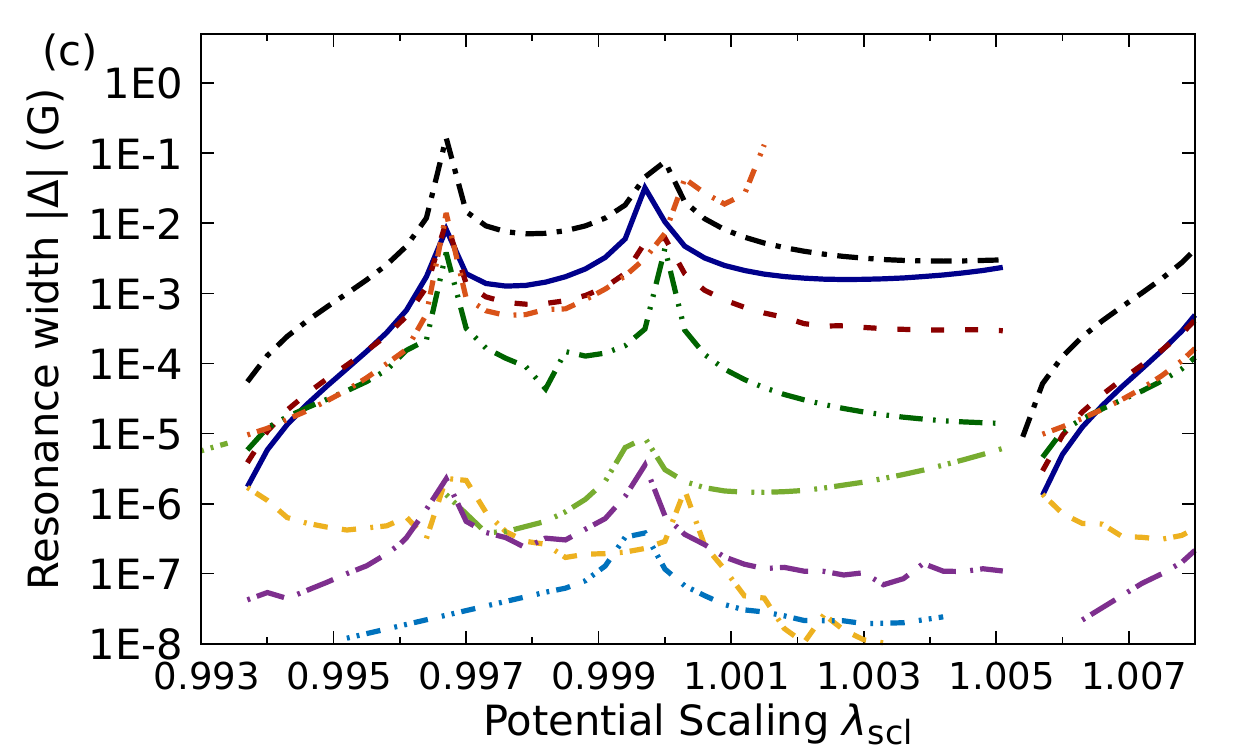}
	}	
    \vspace{-0.5 cm}
\caption{(a) Cube roots of binding energies $E_\textrm{b}$ for states below $f=2$ thresholds, as a function of potential scaling factor; (b) resonance positions $B_\textrm{res}$; (c) resonance widths $|\Delta|$.
}%
\label{fig:res-3P0}
\end{figure}

The resonance positions and widths are sensitive to uncertainties in the interaction potential. A small change in the potential has a large effect on the scattering lengths and binding energies. Figure \ref{fig:res-3P0}(a) shows the calculated binding energies of zero-field bound states for Rb($f=1,m_f=1$) + $^{174}$Yb($^3$P$_0$) for $L=0$ and 2 as a function of the overall scaling $\lambda_\textrm{scl}$ of the interaction potential. It may be seen that the state with vibrational quantum number $n=-4$ relative to the $f=2$ threshold lies just below the $f=1$ threshold when $\lambda_\textrm{scl}$ is slightly greater than 0.993. This same state is labeled $n=-5$ at $\lambda_\textrm{scl}=1$ and above, because another state enters the well just below $\lambda_\textrm{scl}=1$. An increase of 1.2\% in $\lambda_\textrm{scl}$ is sufficient to shift this state from the $f=1$ threshold to 8~GHz below it. At this point a new vibrational state crosses the $f=1$ threshold from above, and the cycle repeats. The states supported by the $f=1$ threshold will show similar cyclic behavior, but the variation in energy is much less because they are much more shallowly bound with respect to the thresholds that support them.

The calculated widths are close to cyclic with the same cycle length, but additional effects operate. In particular they are moderated by the background scattering length $a_\textrm{bg}$ in the incoming channel, which passes through poles and zeroes with the same cycle length of $1.2$\%. The resonance widths are greatly enhanced when $|a_\textrm{bg}|$ is large \cite{Brue:AlkYb:2013}. They are also artificially enhanced near a zero in $a_\textrm{bg}$, though here $a_\textrm{bg} \Delta$ varies smoothly. They are reduced for bound states very close to threshold, which have smaller amplitudes at short range \cite{Brue:AlkYb:2013}, and for resonances at low magnetic fields.

\begin{table}[tbp]
\caption{Calculated parameters for resonances at the threshold Rb($f=1,m_f=1$) + $^{174}$Yb($^3$P$_0$) on the interaction potential with scaling factor $\lambda_\textrm{scl} = 0.996$.}
\begin{footnotesize}
\begin{ruledtabular}
\begin{tabular}{ccccc}
$L_{\rm res},f_{\rm res},m_{f,{\rm res}}$ & $B_{\rm res}$ (G) & $a_{\rm bg}$ ({\AA})  & $\Delta$ (G) & $\bar{\Delta}$ (G)\\
\hline
$0, 2, 1$ & 905 & 13 & $3.8 \times 10^{-3}$ & $1.1 \times 10^{-3}$ \\
$2, 2, 2$ & 622 & 12 & $4.4 \times 10^{-4}$ & $1.2 \times 10^{-4}$ \\
$2, 2, 1$ & 812 & 13 & $3.7 \times 10^{-4}$ & $1.1 \times 10^{-4}$ \\
$2, 2, 0$ & 1140 & 13 & $1.1 \times 10^{-4}$ & $3.3 \times 10^{-5}$ \\
$2, 1, -1$ & 489 & 11 & $2.6 \times 10^{-7}$ & $6.5 \times 10^{-8}$ \\
$2, 1, 0$ & 1162 & 14 & $1.4 \times 10^{-4}$ & $4.5 \times 10^{-5}$ \\
\end{tabular}
\end{ruledtabular}
\label{tab:Delta-bar}
\end{footnotesize}
\end{table}

It is useful to compare the typical strengths of the resonances for Yb($^3$P$_0$) with those for Yb($^1$S) \cite{Brue:AlkYb:2013, Yang:CsYb:2019}. For this we consider the quantity $\bar{\Delta} = a_\textrm{bg} \Delta / \bar{a}$ \cite{Yang:CsYb:2019}, where $\bar{a}$ is the mean scattering length of Gribakin and Flambaum \cite{Gribakin:1993}; this is 43.8 \AA\ for Rb$^{174}$Yb and quite similar for other isotopes. $\bar{\Delta}$ is a better measure of resonance strength than $\Delta$ itself, because it accounts for the artificially large values of $\Delta$ that occur when $a_\textrm{bg}$ is small. Table \ref{tab:Delta-bar} gives values of $B_\textrm{res}$, $a_\textrm{bg}$, $\Delta$ and $\bar{\Delta}$ for $\lambda_\textrm{scl} = 0.996$, which is chosen to give resonances at moderate fields ($B_\textrm{res} \lesssim 1000$~G), with $|a_\textrm{bg}|\lesssim\bar{a}$ to avoid widths enhanced by atypically large values of $|a_\textrm{bg}|$. For resonances due to states with $L_\textrm{res}=0$ supported by the thresholds with $f=2$, we find values of $\bar{\Delta}$ somewhat larger (by up to a factor of 10) than those for resonances at similar fields due to Mechanism I for Rb+Yb($^1$S) \cite{Brue:AlkYb:2013} and Cs+Yb($^1$S) \cite{Yang:CsYb:2019}. Some of the resonances due to states with $L_\textrm{res}=2$ have $\bar{\Delta}$ much larger (by a factor of 10 or more) than those due to Mechanism III for Cs+$^{171}$Yb($^1$S) and Cs+$^{173}$Yb($^1$S) \cite{Yang:CsYb:2019}, and exist for bosonic as well as fermionic isotopes of Yb.

Changing the reduced mass has a very similar effect to scaling the interaction potential by the same fraction. The bosonic isotopes from $^{168}$Yb to $^{176}$Yb, in combination with $^{87}$Rb, offer a set of reduced masses that are approximately equally spaced across a range of 1.6\%. These effectively encompass almost the entire range of behavior shown in Fig.\ \ref{fig:res-3P0}. However, calibration of the interaction potential, using experimental binding energies or scattering lengths, will be needed to predict which specific isotope will produce bound states at a particular depth or resonances at a particular field.

\section{Conclusions}

Magnetically tunable Feshbach resonances exist in ultracold collisions of closed-shell atoms such as Sr ($^1$S) and Yb($^1$S) with alkali-metal atoms, but they are sparse in magnetic field and usually very narrow. Here we have investigated the analogous resonances for Yb atoms in their excited $^3$P$_2$ and $^3$P$_0$ states in collision  with Rb atoms, using coupled-channel scattering and bound-state calculations.

We have obtained spin-free potential-energy curves and spin-orbit coupling functions by fitting to \emph{ab initio} electronic structure calculations based on multireference perturbation theory \cite{Shundalau:2017}. There are 4 spin-free potential curves labelled $^2\Sigma$, $^2\Pi$, $^4\Sigma$ and $^4\Pi$. These are mixed by spin-orbit coupling to produce 9 spin-coupled potential curves. In contrast to previous work, it proved insufficient to use the atomic spin-orbit operator for Yb($^3$P) at all internuclear distances $R$. Instead we need separate $R$-dependent spin-orbit operators involving the spins that are asymptotically on the Yb and Rb atoms.

The $^3$P$_2$ state of Yb lies energetically far above the $^3$P$_1$ and $^3$P$_0$ states. A Yb atom in its $^3$P$_2$ state can therefore undergo inelastic collisions with Rb, even when both atoms are in their lowest Zeeman sublevels. Because of this, Feshbach resonances that exist at Yb($^3$P$_2$) thresholds are decayed, with resonant signatures that show oscillations rather than poles in the scattering length as a function of magnetic field. The molecular states that might be formed by magnetoassociation at these resonances have finite lifetimes. We have explored the resonance structure at several thresholds corresponding to different Zeeman sublevels of Yb($^3$P$_2$) and Rb. The sharpest resonances (with the largest-amplitude variations in scattering length) occur when both atoms are in their lowest sublevels, $m_j=-2$ for Yb and $f=1$, $m_f=1$ for $^{87}$Rb. In some cases the scattering length varies by more than $\pm6000$~\AA\ around the sharp resonances. This contrasts with Li + Yb($^3$P$_2$), where the variations were previously found to be less than $\pm1000$~\AA\ \cite{Gonzalez-Martinez:LiYb:2013}.

The interaction potential we have used is not accurate enough to make absolute predictions of bound-state energies and resonance positions. To account for this, we have considered a range of interaction potentials sufficient to tune the least-bound state in each channel over a complete cycle of possible energies. We have characterized the sharpest resonances observed to obtain both elastic and inelastic widths. The inelastic widths allow us to estimate the lifetimes of the molecular states that could be formed by magnetoassociation, which are at most a few microseconds for RbYb molecules formed at the Yb($^3$P$_2$) thresholds.

We have also investigated Feshbach resonances in collisions of Yb($^3$P$_0$) with Rb. The patterns of the bound states that cause resonances here are closely analogous to those for Yb($^1$S) \cite{Yang:CsYb:2019}. However, there are additional couplings for Yb($^3$P$_0$) that arise directly from the electrostatic potential-energy curves and spin-orbit coupling, and do not rely on the distance-dependence of the hyperfine coupling. These couplings produce resonances due to non-rotating bound states ($L_\textrm{res}=0$) that are typically somewhat stronger (by up to a factor of 10) than the corresponding resonances for Yb($^1$S). These $L$-conserving resonances are quite sparse in magnetic field, as for $^1$S. In addition, there are $L$-changing resonances, due to bound states with $L_\textrm{res}=2$. These are denser as a function of magnetic field. For Yb($^3$P$_0$) they exist even for bosonic (spin-zero) isotopes of Yb; this contrast with Yb($^1$S), where the analogous resonances exist only for fermionic Yb. Some of them are considerably stronger (by a factor of 10 or more) than the $L$-changing resonances that exist for $^{171}$Yb and $^{173}$Yb($^1$S) \cite{Yang:CsYb:2019} and $^{87}$Sr ($^1$S) \cite{Barbe:RbSr:2018}.  Molecules formed by magnetoassociation of Yb($^3$P$_0$) can decay only by processes that form Yb($^1$S), with or without excitation of Rb to its $^2$P state, so are expected to be long-lived.

The data presented in this work are available from Durham University
\cite{DOI_data-RbYb}.

\begin{acknowledgments}
We are grateful to Simon Cornish and Tobias Franzen for valuable discussions. This work was supported by the U.K. Engineering and Physical Sciences Research Council
(EPSRC) Grant EP/P01058X/1.
\end{acknowledgments}

\appendix
\section{Fitted parameters for potential curves and spin-orbit matrix}

The spin-orbit-free potential curves are represented by Hulburt-Hirschfelder potentials~\cite{Hulburt:1941} supplemented with damped dispersion terms,
\begin{eqnarray}
V(R) &=& D_{\rm e}\left[e^{-2x} - 2e^{-x} + px^3e^{-2x}(1 + qx)\right] \nonumber\\
&-& \sum_{n=6,8} D_{n}(\alpha R)C_nR^{-n}.	
\label{eq:pot}
\end{eqnarray}
Here $x=\beta(R-R_{\rm e})$, where $R_{\rm e}$ is the equilibrium distance, and $D_{\rm e}$ is the well depth. Both $R_{\rm e}$ and $D_{\rm e}$ exclude the dispersion terms. The functions $D_n(\alpha R)$ are Tang-Toennies damping functions~\cite{TANG:1984},
\begin{eqnarray}
D_n(\alpha R) = 1 - e^{-\alpha R}\sum_{m=0}^n \frac{(\alpha R)^m}{m!}.
\end{eqnarray}

The dispersion coefficients $C_6$ and $C_8$ are the same for doublet and quartet curves but different for $\Sigma$ and $\Pi$ curves. We obtain the average value $C_6^0 = (1/3)(C_6^{\Sigma}+2C_6^{\Pi}) = 4265.6~E_{\rm h}a_0^6$ for Rb($^2$S) + Yb($^3$P) using Tang's combination rule~\cite{Tang:1969} with the values of the static polarizability and dispersion coefficients for Rb~\cite{Derevianko:2010} and
Yb($^3$P)~\cite{Dzuba:2010}. The difference $C_6^{\Sigma}-C_6^{\Pi}$ is not known for Rb + Yb, so we use the ratio $C_6^{\Sigma}/C_6^{\Pi}=1.146$ that was used for Li($^2$S) + Yb($^3$P)~\cite{Gonzalez-Martinez:LiYb:2013}. This gives $C_6^{\Sigma} = 4661.5$ and $C_6^{\Pi} = 4067.6~E_{\rm h}a_0^6$ for Rb($^2$S) + Yb($^3$P). For each spin-free curve, we use a value of $C_8$  related to $C_6$ by $C_8/C_6 = 80a_0^2$.

From Equations (6) and (9) of Sec. \ref{sec:theory}, we write the total spin-orbit Hamiltonian for Rb($^2$S) + Yb($^3$P) as
\begin{eqnarray}
\hat{H}_{\rm so}(R,\xi) &=& a_{\rm Yb}\hat{l}_{\rm Yb}.\hat{s}_{\rm Yb} + \Delta a_\textrm{Yb}(R) \hat{l}_\textrm{Yb}\cdot\hat{s}_\textrm{Yb} \nonumber\\
&+& \Delta a_\textrm{Rb}(R) \hat{l}_\textrm{Yb}\cdot\hat{s}_\textrm{Rb} + a_1\delta_{j1}.
\label{eq:Hso}
\end{eqnarray}
The functions $\Delta a_{\rm Yb}(R)$ and $\Delta a_{\rm Rb}(R)$ are represented with switching functions,
\begin{eqnarray}
\Delta a_{\rm Yb}(R) &=& -a_{\rm Yb}\epsilon[1 - \tanh\{ \sigma(R-R_0) \}] \label{eq:aso_yb}\\
\Delta a_{\rm Rb}(R) &=& a_{\rm Rb}\epsilon[1 - \tanh\{ \sigma(R-R_0) \}]
\label{eq:aso_rb}
\end{eqnarray}

We have fitted the parameters of Eqs. (\ref{eq:pot}) to (\ref{eq:aso_rb}) to the spin-orbit-coupled \emph{ab initio} potential curves of Shundalau and Minko \cite{Shundalau:2017}, obtained using multireference perturbation theory. At this stage we used atomic spin-orbit coupling constants chosen to match the \emph{ab initio} curves, $a_{\rm Yb}=807$ cm$^{-1}$ and $a_1=0$ \footnote{We were able to reproduce most of the curves at most distances, but at some distances the electronic structure calculations for $\Omega=1/2$ appear to have followed an ``interloper state'' instead of one of the curves for Rb ($^2$S) + Yb ($^3$P).}.
The fitted parameters are given in Tables \ref{tab:potparam} and \ref{tab:soparam}.

\begin{table}[htp]
\caption{Parameters of the spin-free interaction potentials.}
\begin{ruledtabular}
\begin{tabular}{ccccccc}
$V$ & $D_{\rm e}$ (cm$^{-1}$) & $R_{\rm e}$ ({\AA}) &  $\beta$ ({\AA}$^{-1}$) & $p$ & $q$ & $\alpha$ ({\AA}$^{-1}$)\\
\hline
$^2\Pi$ & 4460.00 & 4.45 & 0.64 & 0.38 & 0.80 & 0.70 \\
$^2\Sigma$ & 3500.00 & 4.90 & 0.61 & 0.40 & 0.50 & 0.70 \\
$^4\Pi$ & 1812.12 & 4.47 & 0.89 & 0.20 & 0.84 & 0.70 \\
$^4\Sigma$ & 157.79 & 6.75 & 0.69 & 0.15 & 0.49 & 0.90 \\
\end{tabular}
\end{ruledtabular}
\label{tab:potparam}
\end{table}

\begin{table}[htp]
\caption{Parameters of the spin-orbit coupling.}
\begin{ruledtabular}
\begin{tabular}{cccccc}
$a_{\rm Yb}$ (cm$^{-1}$) & $a_{\rm Rb}$ (cm$^{-1}$) & $\epsilon$ &  $\sigma$ ({\AA}$^{-1}$) & $R_0$ ({\AA})
& $a_1$ (cm$^{-1}$)\\
\hline
806.612 & 524.283 & 0.20 & 0.88 & 6.66 & 0
\end{tabular}
\end{ruledtabular}
\label{tab:soparam}
\end{table}

In our coupled-channel calculations, we replace the values of $a_{\rm Yb}$ and $a_1$ from the \emph{ab initio} calculations with those that reproduce the experimental splittings of Yb($^3$P), $a_{\rm Yb} = 807.3163$ and $a_1 = -103.7483$ cm$^{-1}$.

\section{Matrix elements of $\boldsymbol{\hat{l}_{\rm Yb}.\hat{s}_{\rm Rb}}$}

The matrix elements of the operator $\hat{l}_{\rm Yb}.\hat{s}_{\rm Rb}$ in the basis set
$|s_{\rm Rb}m_{s,{\rm Rb}}\rangle |i_{\rm Rb}m_{i,{\rm Rb}}\rangle |(l_{\rm Yb}s_{\rm Yb})jm_j\rangle  |LM_L\rangle$ are diagonal in $L$, $M_L$,
$i_{\rm Rb}$ and $m_{i,{\rm Rb}}$. The remaining factors are
\begin{widetext}
\begin{eqnarray}
&\,& \langle (l_{\rm Yb}s_{\rm Yb}) jm_j |
\langle s_{\rm Rb} m_{s,{\rm Rb}} |
\hat{l}_{\rm Yb}.\hat{s}_{\rm Rb}
|s_{\rm Rb} m_{s,{\rm Rb}}^{\prime} \rangle
| (l_{\rm Yb}s_{\rm Yb})j^{\prime} m_j^{\prime}\rangle
\nonumber \\
&=& (-1)^{(1+s_{\rm Yb}+l_{\rm Yb}+j+3j^{\prime})}
[s_{\rm Rb}(s_{\rm Rb}+1)(2s_{\rm Rb} + 1)]^{\frac{1}{2}} [l_{\rm Yb}(l_{\rm Yb}+1)(2l_{\rm Yb} + 1)]^{\frac{1}{2}}
[(2j + 1)(2j^{\prime}+1)]^{\frac{1}{2}}
\begin{Bmatrix}
j & j^{\prime} & 1 \\
l_{\rm Yb} & l_{\rm Yb} & s_{\rm Yb}
\end{Bmatrix}
\nonumber \\
& \times &
\sum_{J_{\rm m}} (-1)^{s_{\rm Rb}+J_{\rm m}}(2J_{\rm m} + 1)
\begin{pmatrix}
j & s_{\rm Rb} & J_{\rm m}  \\
m_j & m_{s,{\rm Rb}} & -M_{J,{\rm m}}
\end{pmatrix}
\begin{pmatrix}
j^{\prime} & s_{\rm Rb} & J_{\rm m}  \\
m_j^{\prime} & m_{s,{\rm Rb}}^{\prime} & -M_{J,{\rm m}}
\end{pmatrix}
\begin{Bmatrix}
j & j^{\prime} & 1 \\
s_{\rm Rb} & s_{\rm Rb} & J_{\rm m}
\end{Bmatrix}.
\end{eqnarray}
\end{widetext}
Here (:::) and \{:::\} are Wigner 3-$j$ and 6-$j$ symbols. $J_{\rm m}$ is the resultant of $j$ and $s_{\rm Rb}$, so the three of these must satisfy triangle conditions, and $M_{J,{\rm m}} = m_j + m_{s,{\rm Rb}} =
m_j^{\prime} + m_{s,{\rm Rb}}^{\prime}$.

\bibliographystyle{long_bib}
%\bibliography{../CsYb_resonances/Resubmission/CsYb_res,../all}
\bibliography{../all,RbYb3PData}
\end{document}